\documentclass[letterpaper]{aa520}  
\usepackage{graphicx}
\usepackage{txfonts}
\usepackage{natbib}

\newcommand{\HI}{\ion{H}{i}}
\newcommand{\HII}{\ion{H}{ii}}
\newcommand{\HeI}{\ion{He}{i}}
\newcommand{\HeII}{\ion{He}{ii}}
\newcommand{\FeII}{[\ion{Fe}{ii}]}
\newcommand{\FeIII}{[\ion{Fe}{iii}]}
\newcommand{\NiqII}{[\ion{Ni}{ii}]}
\newcommand{\NeII}{[\ion{Ne}{ii}]}
\newcommand{\OI}{\ion{O}{i}}
\newcommand{\fOI}{[\ion{O}{i}]}
\newcommand{\OII}{[\ion{O}{ii}]}
\newcommand{\OIII}{[\ion{O}{iii}]}
\newcommand{\CII}{[\ion{C}{ii}]}

\newcommand{\NII}{[\ion{N}{ii}]}
\newcommand{\NIII}{[\ion{N}{iii}]}
\newcommand{\SilII}{\ion{Si}{ii}}
\newcommand{\NaI}{\ion{Na}{i}}
\newcommand{\KI}{\ion{K}{i}}
\newcommand{\um}{$\mu$m}
\newcommand{\ClIII}{[\ion{Cl}{iii}]}
\newcommand{\ArIII}{[\ion{Ar}{iii}]}
\newcommand{\SII}{[\ion{S}{ii}]}
\newcommand{\SIII}{[\ion{S}{iii}]}
\newcommand{\NI}{\ion{N}{i}}

\begin{document}
   \title{M1--78: a nitrogen-rich Galactic compact \HII\ region beyond the Perseus arm}

   \author{N.\,L. Mart\'{\i}n-Hern\'{a}ndez
          \inst{1}
	  \and
          C. Esteban
	  \inst{1}
	  \and
	  A. Mesa-Delgado
	  \inst{1}
	  \and
	  A. Bik
	  \inst{2}
	  \and
	  E. Puga
	  \inst{3}
          }

   \offprints{N.L.\,Mart\'{\i}n-Hern\'{a}ndez, leticia@iac.es}

   \institute{Instituto de Astronom\'ia de Canarias, V\'ia L\'actea s/n,
             38205 La Laguna, Spain
             \and
	     European Southern Observatory, Karl-Schwarzschild Strasse 2, 85748 Garching-bei-M\"{u}nchen, Germany
	     \and
	     Instituut voor Sterrenkunde, Katholieke Universiteit Leuven, Celestijnenlaan 200B, 3001 Leuven, Belgium
	     }

   \date{Received; accepted}

 
  \abstract
   {There is  considerable controversy surrounding the nature of M1--78, a compact nebula 
located beyond the Perseus arm. It was first classified as a 
planetary nebula and is nowadays generally considered to be a compact \HII\ region.} 
   {To investigate the nature M1--78 further, we present a detailed spectroscopic study of M1--78 in the optical and near-infrared.}
   {We obtained long-slit, 
intermediate-resolution, optical spectroscopy with the ISIS spectrograph mounted on the William 
Herschel Telescope (WHT) at Roque de los Muchachos Observatory (La Palma, Spain). 
As a complement, we obtained long-slit, intermediate-resolution, near-infrared
spectra using LIRIS, the near-infrared imager/spectrographer also installed at the WHT.}
   {M1--78 is a high-density nebula with substantial physical 
differences between its two main morphological zones: a bright arc to the SW 
and a blob of emission in the NE. Specifically, the blob in the NE has 
a higher electron temperature (13400~K) and visual extinction (about 9~mag) than the 
SW arc. The most important result, however, is 
the confirmation of a nitrogen enrichment in M1--78. This enrichment 
is stronger at the location of the NE blob and is correlated with a defficiency 
in the O abundance and a (dubious) He enrichment. Such an abundance pattern is typical 
of ejecta nebulae around evolved massive stars such as Wolf--Rayet and 
Luminous Blue Variable stars. The spatial variations in the physical conditions 
and chemical abundances and the presence of more than one possible ionizing source 
indicates, however, that M1--78 is better described as a combination of a compact \HII\ 
region $+$ ejecta. This is confirmed by the  \HeI\,2.112\,\um/Br$\gamma$ line ratio, 
which indicates a hot ($T_{eff} \gtrsim 40000$~K) O star in the SW arc. 
Finally, we detect H$_2$ emission that extends over a large ($\sim30$\arcsec) area around 
the ionized nebula. Analysis of the near-infrared H$_2$ lines indicates that the excitation
 mechanism is UV fluorescence.} 
     {}  
  

   \maketitle
   
%

\section{Introduction}

M1--78 (another designation is IRAS\,21190$+$5140) was first observed by 
\cite{minkowski46} in a survey of emission-line objects and is nowadays 
generally classified as an \HII\ region. Nevertheless, it is still listed 
in the {\it Simbad Astronomical Database}\footnote{http://simbad.u-strasbg.fr/simbad/} 
as a ``possible planetary nebula'' (PN), which reflects the controvery 
that has existed around the nature of this source. The survey of 
\cite{latter95} catalogues this object as a middle-butterfly PN, where 
a butterfly or bipolar nebula is a PN that is bi-lobed and has an
 equatorial waist. \cite{phillips04} places this source at a distance 
of 1.6~kpc using the correlation between 5~GHz brightness temperatures 
and intrinsic radio luminosities established for PNe and based on stellar 
evolutionary models. However, the distance determined by \cite{puche88} 
based on the \HI\ absorption spectrum is in between 5 and 8~kpc, beyond 
both the Local arm and the Perseus arm. This is in agreement with the
 kinematic distance of 8.9~kpc obtained from its radial velocity 
\citep[e.g.][]{peeters:catalogue}. Such a distance implies a high 
luminosity of about $2\times10^5$~L/L$_\odot$ 
\citep[see again][]{peeters:catalogue}, which makes the classification 
of M1--78 as a PN untenable, as pointed out already by \cite{zijlstra90} 
and \cite{gussie95}. A classification as an ultracompact \HII\ region 
is favored instead by these authors. \cite{gussie95} even favors a 
rather unique classification for this region, that of an ultracompact 
\HII\ region with a post-main sequence central star, probably a Wolf-Rayet (WR) star, 
based on the high expansion velocity of the nebula 
(around 25 km\,s$^{-1}$) and a possible overabundance of nitrogen 
\citep[see also][]{perinotto91}.

Several near-infrared (NIR) spectra of M1--78 have been obtained showing emission of 
atomic hydrogen and helium and molecular hydrogen 
\citep[e.g.][]{hora99,lumsden01,lumsden01b}. This source has also been 
imaged in the NIR \citep[e.g.][]{latter95}, mid-infrared 
\citep[e.g.][]{peeters:thesis} and radio \citep[e.g.][]{zijlstra90}, 
showing a similar morphology at all wavelengths, i.e. an overall 
structure emcompassing a prominent arc of which the tail at both 
sides converges towards a blob of emission to the NE 
(see Fig.~\ref{fig:im}). The ISO spectrum of this object is rather
 unique and shows silicate in emission at about 10 and 20~\um\ 
\citep{peeters:catalogue}. In addition, it shows strong emission 
features generally attibuted to emission by polyclyclic aromatic 
hydrocarbon (PAH) molecules, typical of \HII\ regions 
\citep[cf.][]{peeters:catalogue}. 

We obtained long-slit, intermediate-resolution, optical 
spectroscopy with the ISIS spectrograph mounted on the 4.2\,m William 
Herschel Telescope (WHT) at Roque de los Muchachos Observatory 
(La Palma, Spain). As a complement, we obtained long-slit, 
intermediate-resolution, NIR spectra of M1--78 using LIRIS, the NIR 
imager/spectrographer also mounted on the WHT. The aim of this 
study is to investigate the controversial nature of M1--78.

This paper is structured as follows. Section~\ref{sect:observations} 
describes the observations and data reduction. 
Section~\ref{sect:results:opt} and Sect.~\ref{sect:results:nir} present 
the results of the optical and NIR data, respectively. 
Section~\ref{sect:discussion} discusses the nature of M1--78 and 
 Sect.~\ref{sect:conclusions} summarizes the main conclusions.

\section{Observations and data reduction}
\label{sect:observations}

\subsection{Optical observations}
\label{obs:optical}

Intermediate-resolution spectroscopy was obtained on 2007 July 18 with the ISIS spectrograph  
mounted on the 4.2\,m William Herschel Telescope (WHT) at Roque de los Muchachos Observatory 
(La Palma, Spain). Two different CCDs were used at the blue and red arms of the spectrograph: 
an EEV CCD with a configuration 4100 $\times$ 2048 pixels with a pixel size of 13.5 $\mu$m in the blue arm and a 
REDPLUS CCD with 4096 $\times$ 2048 pixels with a pixel size of 15 $\mu$m in the red arm. The dichroic used to separate 
the blue and red beams was set at 5300 \AA. The slit was 3\farcm7 long and 1\farcs03 wide. Two gratings 
were used, the R1200B in the blue arm and the R316R in the red arm. These gratings give reciprocal 
dispersions of 17 and 62 \AA\ mm$^{-1}$, and effective spectral resolutions of 0.86 and 3.56 \AA\ 
for the blue and red arms, respectively. The blue spectra cover from $\lambda\lambda$4225 to 5075 \AA\ and the red 
ones from $\lambda\lambda$5335 to 8090 \AA. The spatial scale is 0\farcs20 pixel$^{-1}$ and 0\farcs22 pixel$^{-1}$ 
in the blue and 
red arms, respectively. The average seeing during the observations was $\sim$0\farcs8.

\begin{figure}
   \centering \resizebox{\hsize}{!}{\includegraphics{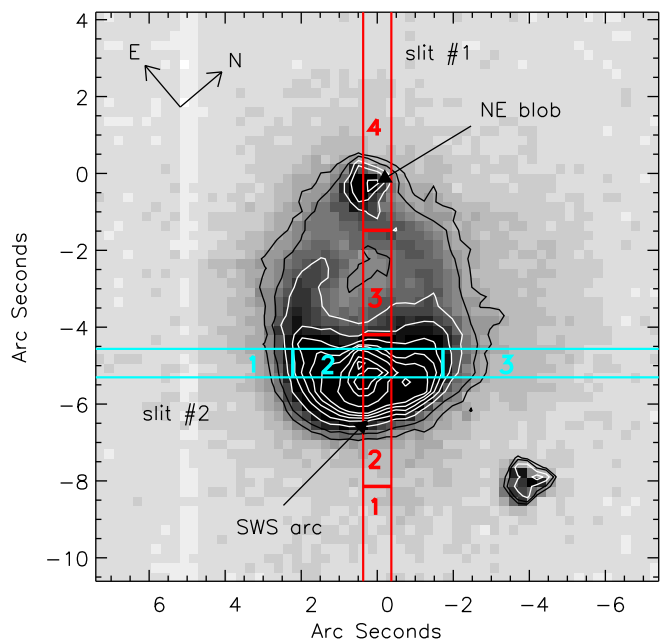}}
     \caption{NIR broad-band ($Ks$ filter) acquisition image of M1--78.
     The positions of the NIR slits (in red slit \#1 with a P.A. of 50\degr, and in cyan slit \#2 
     with a P.A. of 140\degr) are marked. The (0\arcsec,0\arcsec) position corresponds to 
     R.A.=21$^h$20$^m$44.91$^s$ and Dec.=$+51^d53^m27.3^s$ (J2000.0). The H$_2$ regions 
analysed in Sect.~\ref{sect:h2} are indicated. At a distance of 8.9~kpc, 1\arcsec\ corresponds 
       to 0.043~pc.}
        \label{fig:im}
   \end{figure}

We observed a single slit position centered on the brightest nebular spot
 (see Fig.~\ref{fig:im}) and with a position angle P.A.=50\degr.
We took three consecutive 600\,s exposures in both arms in order to get a reasonably good spectrum of this faint object. 
The data were wavelength calibrated with a CuNe+CuAr lamp. The correction 
for atmospheric extinction was performed using the 
average curve for continuous atmospheric extinction at Roque de los Muchachos Observatory. The absolute flux calibration 
was achieved by observations of the standard stars BD+33~2642, BD+28~4211 and BD+25~4655. 
All the CCD frames were reduced using 
the standard IRAF\footnote{IRAF is distributed by the National Optical Astronomy Observatory,
 which is operated by the Association of Universities for the Research in Astronomy, Inc., under 
cooperative agreement with the National Science Foundation (http://iraf.noao.edu/).} TWODSPEC
 reduction package to perform 
bias correction, flat-fielding, cosmic-ray rejection, wavelength and flux calibration, and sky subtraction.

\subsection{NIR observations}
\label{obs:nir}

NIR spectra in the 2.0--2.4~\um\ wavelength range ($K$-band) were obtained on the night of 
2006 September 9 at the 4.2\,m WHT using LIRIS, a long-slit, intermediate-resolution, infrared 
spectrograph \citep{acosta03,manchado04}. The spectral region blueward of about 1.98\,\um\ is 
affected by substantially degraded atmospheric transmission.
LIRIS is equipped with a Rockwell Hawaii $1024\times1024$ HgCdTe array detector. The spatial 
scale is 0\farcs25 pixel$^{-1}$, and the slit width used during the observations was 0\farcs75, 
allowing a spectral resolution $R=\lambda/\Delta\lambda\sim 1000$. The slit was orientated along 
P.A.$=50\degr$ (slit \#1, which corresponds to the single-slit observed in the optical) and 
140\degr\ (slit \#2) centered on the brightest nebular spot (see Fig.~\ref{fig:im}). Weather 
conditions were relatively good, although with sparse cirrus. The seeing during our 
observations varied between 0\farcs8 and 1\farcs3 measured from the FWHM of the standard star spectra.

Observations were performed following an ABBA telescope-nodding pattern, placing the source in two 
positions along the slit, separated by 45\arcsec. Individual frames were taken with integration times 
of 300\,s and total on-source integration times of 30 minutes for slit \#1 and 20 minutes for 
slit \#2. In order to obtain the telluric correction and the flux calibration, the nearby 
HIP\,106393 A0\,V star was observed with the same configuration. The data were reduced following 
standard procedures for NIR spectroscopy, using IRAF and LIRISDR, the LIRIS Data Reduction package. 
After the flat-field correction, consecutive pairs of AB two-dimensional spectra were subtracted to 
remove the sky background. The resulting frames were then co-added to provide the final spectrum. 
Sky lines present in the science data were used to determine the wavelength calibration. The 
final uncertainty in the wavelength calibration is roughly 1\AA. The resulting
 wavelength-calibrated spectra were divided by a composite spectrum to remove the telluric 
contamination. This composite spectrum was generated from the observed spectra of HIP\,106393, 
divided by a Vega model convolved with the actual spectral resolution ($\sim 23$\AA) using the 
routine {\it xtellcor\_general} within the Spextool package\footnote{http://irtfweb.ifa.hawaii.edu/$\backsim$spex/}, 
an IDL-based spectral reduction tool written by Mike Cushing and Bill Vacca. This routine 
is also used for the flux calibration, which is done by normalizing with the B and V magnitudes 
of the standard star provided by the {\it Simbad Astronomical Database} (B=5.734 and V=5.766).
 We estimate an uncertainty in the flux calibration around 40\% based on the comparison of the 
various spectra obtained for the standard star.

\section{Results from optical observations}
\label{sect:results:opt}

\subsection{Features of the optical spectra}

\begin{figure}
   \centering \resizebox{\hsize}{!}{\includegraphics{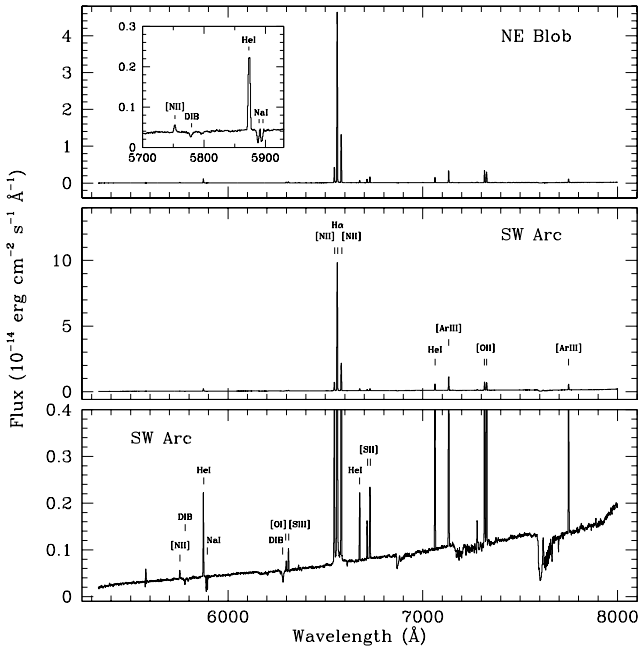}}
     \caption{Red range of the optical spectra extracted from a slit position 
coincident with slit \#1 of the NIR observations (see Fig.~\ref{fig:im}). 
The upper panel corresponds to the emission of the NE blob; the inset 
show an enlargment of the spectral area around the \NII\ 5755 \AA\ and 
\HeI\, 5876\,\AA\ lines. The middle and the lower panels show the red spectrum 
of the SW arc at different line-flux levels. The fluxes are not 
corrected for reddening.  The identification of most emission lines and 
absorption features are indicated. The broad absorption features longward 
6800 \AA\ are due to the atmosphere.}
        \label{fig:sp:optical}
   \end{figure}

The two-dimensional optical spectrum shows two peaks in the surface brightness 
distribution of the emission lines along the spatial direction. These peaks coincide 
with the two main morphological zones of the nebula: the brightest arc to the 
SW and the blob of emission to the NE. In addition, each peak 
contains a featureless, very weak and unresolved stellar continuum  (note that 
the nebular contribution to the continuum in the optical is generally very small). 
Only the continuum associated with the SW arc  -- brightest one -- is 
detected in the spectrum of the blue arm. We extracted two one-dimensional spectra 
isolating the emission from each peak, the spatial extension of each  zone being
4\farcs4. The spectra of the red range of both zones are shown in 
Fig.~\ref{fig:sp:optical}. A detailed inspection of the emission line profiles 
along the slit does not show any evidence of 
stellar features or broad components in the recombination-line profiles due to
 the contribution of photospheric emission of massive stars.

Line intensities were measured integrating all the flux in the line between two given limits and over a local continuum 
estimated by eye and using the SPLOT routine of the IRAF package. 
The line intensities of the blue arm spectra were normalized to that of H$\beta$, while those of the red arm 
were normalized to H$\alpha$. The reddening coefficient, $c$(H$\beta$), was determined by fitting the 
observed $I$(H$\gamma$)/$I$(H$\beta$) ratio of the spectra of the blue arm to the theoretical line ratio obtained from the  
calculations of \cite{storey95}, assuming case B,  
and electron temperature and density of 10$^4$ K and $n_e$ = 10$^3$ cm$^{-3}$. 
We considered the reddening law of \cite{whitford58}. 
Once the reddening was derived, the emission line ratios of the red spectra were re-scaled to the $I$(H$\beta$). 
The spectrum of the NE blob shows a somewhat larger reddening coefficient, 
indicating a larger amount of internal dust in this 
zone of M1--78 that produces an extinction A$_V$ of about 9 mag. Our determinations 
of $c$(H$\beta$) are consistent with the value of 4.1 
found by \cite{aller87}.  
 \cite{lumsden01b} derived a similar extinction in the NIR of A$_V \sim 6.5$~mag 
\citep[assuming $R_V=3$;][]{mathis90} estimated by comparing the higher Brackett 
series lines with Br$\gamma$ and assuming an exponential extinction law $A_\lambda \propto \lambda^{-1.85}$.

\begin{table*}
\caption{Dereddened line intensity ratios with respect to $I$(H$\beta$) =100,  
reddening coeeficient, observed H$\beta$ flux and physical conditions obtained from the optical spectra.}
\label{table:linesoptical}
\begin{center}
\begin{tabular}{lcccccc}
\hline\hline\\[-5pt]
& & & \multicolumn{2}{c}{SW arc} & \multicolumn{2}{c}{NE blob} \\ \cline{4-5} \cline{6-7} \\[-5pt]  
$\lambda_{\rm 0}(\AA)$ & Identification & $f$($\lambda$) & $\lambda_{\rm obs}(\AA)$ & $I$($\lambda$)/$I$(H$\beta$) & 
$\lambda_{\rm obs}(\AA)$ & $I$($\lambda$)/$I$(H$\beta$) \\
\hline\\[-5pt]
  4340.47 & H$\gamma$ & 0.135 & 4339.00 & 47.0 $\pm$ 2.5 & 4339.04 & 47.0 $\pm$ 3.9 \\
  4471.48 & \HeI & 0.10 & 4470.09 & 4.0 $\pm$ 0.8 & ... & ... \\
  4861.33 & H$\beta$ & 0.00 & 4859.73 & 100 $\pm$ 2 & 4859.77 & 100 $\pm$ 2 \\
  4921.93 & \HeI & -0.01 & 4920.42 & 9.2: & ... & ... \\
  4958.91 & \OIII & -0.02 & 4957.30 & 134.6 $\pm$ 3.8 & 4957.38 & 93.6 $\pm$ 2.7 \\
  5006.84 & \OIII & -0.03 & 5005.17 & 372.1 $\pm$ 8.3 & 5005.26 & 258.3 $\pm$ 5.8 \\
  5015.68 & \HeI & -0.03 & 5013.99 & 2.1 $\pm$ 0.4 & 5014.15 & 2.0 $\pm$ 0.4 \\
  5517.71 & \ClIII & -0.17 & ... & ... & 5516.19 & 0.8 $\pm$ 0.2 \\
  5537.88 & \ClIII & -0.18 & ... & ... & 5536.28 & 1.0 $\pm$ 0.2 \\
  5754.64 & \NII & -0.21 & 5752.77 & 1.3 $\pm$ 0.2 & 5752.69 & 2.5 $\pm$ 0.1 \\
  5875.67 & \HeI & -0.23 & 5873.75 & 11.4 $\pm$ 0.4 & 5873.72 & 19.8 $\pm$ 0.6 \\ 
  6046.44 & \OI & -0.26 & ... & ... & 6044.87 & 2.5: \\
  6300.30 & \fOI & -0.30 & 6298.55 & 0.9 $\pm$ 0.1 & 6298.50 & 2.4 $\pm$ 0.1 \\
  6312.10 & \SIII & -0.30 & 6310.38 & 1.8 $\pm$ 0.1 & 6310.27 & 2.4 $\pm$ 0.1 \\
  6363.78 & \fOI & -0.31 & 6362.32 & 0.2: & 6361.96 & 0.7 $\pm$ 0.1 \\
  6571.36 & \SilII & -0.31 & ... & ... & 6370.07 & 2.9 0.7 $\pm$ 0.4 \\
  6548.03 & \NII & -0.34 & 6546.10 & 18.5 $\pm$ 0.5 & 6546.02 & 26.4 $\pm$ 0.8 \\ 
  6562.82 & H$\alpha$ & -0.34 & 6560.85 & 286.0 $\pm$ 5.9 & 6560.74 & 286.0 $\pm$ 5.9 \\
  6583.41 & \NII & -0.34 & 6581.41 & 59.4 $\pm$ 1.3 & 6581.32 & 59.4 $\pm$ 1.3 \\
  6678.15 & \HeI & -0.35 & 6676.08 & 3.8 $\pm$ 0.1 & 6676.00 & 3.8 $\pm$ 0.1 \\
  6716.47 & \SII & -0.36 & 6714.33 & 2.1 $\pm$ 0.1 & 6714.05 & 2.1 $\pm$ 0.1 \\
  6730.85 & \SII & -0.36 & 6728.76 & 3.8 $\pm$ 0.1 & 6728.46 & 3.8 $\pm$ 0.1 \\
  7001.92 & \OI & -0.39 & ... & ... & 6999.58 & 0.2: \\
  7065.28 & \HeI & -0.40 & 7062.74 & 8.8 $\pm$ 0.3 & 7062.70 & 8.8 $\pm$ 0.3 \\
  7135.78 & \ArIII & -0.41 & 7133.23 & 17.4 $\pm$ 0.5 & 7133.11 & 17.4 $\pm$ 0.5 \\
  7155.14 & \FeII & -0.41 & ... & ... & 7152.25 & 0.1: \\
  7254.35$^{(a)}$ & \OI & -0.42 & ... & ... & 7251.66 & 2.3 $\pm$ 0.8 \\
  7281.35 & \HeI & -0.43 & 7279.25 & 1.0 $\pm$ 0.2 & 7278.96 & 0.43 $\pm$ 0.04 \\
  7318.39 & \OII & -0.43 & 7317.43 & 9.8 $\pm$ 0.4 & 7317.18 & 8.7 $\pm$ 0.3 \\
  7329.66 & \OII & -0.43 & 7327.83 & 9.1 $\pm$ 0.3 & 7327.59 & 7.5 $\pm$ 0.2 \\
  7377.83 & \NiqII & -0.44 & 7375.68 & 0.2: & 7374.98 & 0.20 $\pm$ 0.02 \\
  7442.30 & \NI & -0.44 & ... & ... & 7438.75 & 0.09 $\pm$ 0.02 \\
  7468.31 & \NI & -0.45 & ... & ... & 7465.74 & 0.15 $\pm$ 0.03 \\
  7751.10 & \ArIII & -0.48 & 7748.92 & 4.5 $\pm$ 0.1 & 7748.79 & 1.56 $\pm$ 0.06 \\
 \hline\\[-5pt]
  $c$(H$\beta$) & & & \multicolumn{2}{c}{3.09 $\pm$ 0.02} & \multicolumn{2}{c}{4.10 $\pm$ 0.04} \\
  $F$(H$\beta$)$^{(b)}$ & & & \multicolumn{2}{c}{160 $\pm$ 3} & \multicolumn{2}{c}{73 $\pm$ 2} \\
  $n_{\rm e}$(\SII)$^{(c)}$ & & & \multicolumn{2}{c}{6700 $\pm$ 1000} & \multicolumn{2}{c}{6400 $\pm$ 1200} \\
  $n_{\rm e}$(\ClIII)$^{(c)}$ & & & \multicolumn{2}{c}{...} & \multicolumn{2}{c}{5400$^{+8800}_{-900}$} \\
  $T_{\rm e}$(\NII)$^{(d)}$ & & & \multicolumn{2}{c}{10900 $\pm$ 600} & \multicolumn{2}{c}{13400 $\pm$ 400} \\
\hline
\end{tabular} 
\end{center}
$^{(a)}$ Blend of \OI\ lines at 7254.14, 7254.45 and 7254.53 \AA.
$^{(b)}$ Observed (uncorrected) flux in units of 10$^-16$ erg cm$^{-2}$ s$^{-1}$.
$^{(c)}$ In cm$^{-3}$.
$^{(d)}$ In K.
\end{table*}

In Table~\ref{table:linesoptical} we include, for the two extracted areas, the 
observed and laboratory wavelength of the identified lines, their reddening-corrected
intensity ratios with respect to H$\beta$, the corresponding values of the reddening function, $f$($\lambda$), 
for each emision line, the reddening coefficient $c$(H$\beta$), the observed 
H$\beta$ line flux and the physical conditions of the gas  -- electron density and 
temperature -- determined from different emission-line ratios. We performed a 
single Gaussian fit of the profiles of the brightest lines of the blue range of 
the spectra (H$\beta$ and \OIII\,4959 and 5007\,\AA) for determining the radial 
velocity of the nebula with respect to the local standard of rest (LSR), finding 
a value of $-$74$\pm$1 and $-$70$\pm$2 km s$^{-1}$ for the SW arc and the 
NE blob, respectively. These values are in 
good agreement with the 
v$_{\rm LSR}$ = $-$76$\pm$2 km s$^{-1}$ found by \cite{gussie89} also from optical 
spectroscopic observations and with measurements of hydrogen radio 
recombination lines obtained by \cite{terzian74} and \cite{churchwell76}. 

The red spectrum of the NE blob shows several uncommon permitted faint 
emission lines that were identified as \OI\ 7254 \AA\ (a blend of three 
unresolved lines at 7254.14, 7254.45 and 7254.53\,\AA), \OI\ 7002 \AA\ 
(a blend of 7001.92 and 7002.23\,\AA) and \OI\ 6046 \AA\ (a blend of 6046.23, 
6046.44 and 6046.49 \AA), belonging to multiplets no. 20, 21 and 22, 
respectively; \NI\ 7442 and 7468 \AA\ of multiplet no. 3; and \SilII\ 6347 and 6371 \AA\ of multiplet no. 3. As 
 demonstrated by \cite{grandi75,grandi76}, all these lines come from 
upper levels of their respective ions that can be excited from the ground state by 
the absorption of stellar photons longward of 912 \AA. Also, it is important to 
note that all these lines are precisely the brightest permitted ones 
of \OI, \NI\ and \SilII\ in the Orion Nebula \citep[see][]{esteban04}, although
 their line ratios with respect to H$\beta$ are larger in M1--78
than in the Orion Nebula. These permitted lines are produced in the vicinity of 
the transition zone of the nebula, where the fraction of neutral material becomes 
important and are rather uncommnon in PN \citep[e.g.][]{liu04},
 where they should be presumably produced in dense clumps or globules 
of neutral material embedded inside the ionized gas \citep{grandi76}. 
Other interesting lines in the spectra of M1--78 are those of 
\FeII\ 7155 \AA\ (multiplet 14F, only detected in the NE blob) 
and \NiqII\ 7378 \AA\ (multiplet 2F, detected in both spectra). Both are also 
the brightest forbidden lines of those ions in the Orion Nebula in the 
spectral range covered by our red arm spectrum \citep{esteban04}. 
\cite{lucy95} and \cite{bautista96} demonstrate that, in the Orion 
Nebula, these lines are produced by the absorption of stellar non-ionizing 
UV photons. Interestingly, these lines, especially those of \NiqII, 
are very strong in the circumstellar ejecta around the Luminous Blue 
Variable (LBV) star P Cygni \citep{johnson92} and another two nebulosities associated 
with LBV stars in the LMC \citep{stahl86}. Stellar fluorescence also seems 
to be the most likely excitation mechanism of \NiqII\ lines in those nebulae \citep{lucy95}.    

As it can be seen in Fig.~\ref{fig:sp:optical}, both one-dimensional 
spectra show clear absorption features. Although the most conspicuous 
ones are atmospheric bands, the narrower features are of interstellar 
origin. We identified the \NaI\ doublet at 5890 and 5896 \AA, the 
\KI\ doublet at 7665 and 7699 \AA, and several diffuse interstellar bands 
(DIBs) at 5780, 5797, 6177, 6284, 6379 and 6614 \AA. We determined the LSR velocities of 
the centroids of the single Gaussian fits of the interstellar absorption 
features comparing the observed and rest frame wavelengths \citep[for the DIBs, we 
used the precise wavelengths of][]{galazutdinov00}. The 
$v_{\rm LSR}$ values we found are in the range $-$68 to $-$32 km\,s$^{-1}$, 
consistent with the velocities of the \HI\ absorption features observed by 
\cite{puche88} and identified as caused by neutral gas in and beyond the 
Perseus spiral arm. The relatively high negative radial velocity of the
 nebula and the presence of these interstellar absorption systems in the 
line of sight of the object clearly indicate that M1--78 is a rather distant Galactic object.

\subsection{Physical conditions and chemical abundances}
\label{sect:opresults}

The physical conditions of the nebula, electron density and temperature 
were derived with the IRAF task TEMDEN of the package NEBULAR, 
based in the FIVEL program developed by \cite{derobertis87} and improved 
by \cite{shaw95}. The electron density, $n_{\rm e}$, was 
calculated from the \SII\ 6717/6731 line ratio in the two one-dimensional spectra as well as the 
\ClIII\ 5518/5538 line ratio in the spectrum of the NE blob. The densities found are very high and similar in  
both zones, 6700 and 
6400 cm$^{-3}$ for the arc and the blob, respectively 
(see Table~\ref{table:linesoptical}), and are roughly consistent but slightly higher than  previous determinations  
based on optical \citep[$\sim 2500$~cm$^{-3}$, ][]{aller87} or far-infrared 
spectroscopy \citep[$\sim 3000$~cm$^{-3}$,][]{martin:paperii}. These large values 
of $n_{\rm e}$ are common in compact \HII\ regions and are also on the order
 of those observed in the Orion Nebula \citep[e.g.][]{esteban98}.

\begin{table}
\caption{Ionic and total abundances derived from optical lines.}
\label{table:abundances}
\begin{center}
\begin{tabular}{lcc}
\hline\hline\\[-5pt]
& SW arc & NE blob  \\
\hline\\[-5pt]
  12+log(O$^+$/H$^+$) & 7.81 $\pm$ 0.17 & 7.31 $\pm$ 0.10 \\
  12+log(O$^{++}$/H$^+$) & 8.00 $\pm$ 0.09 & 7.59 $\pm$ 0.05 \\    
  12+log(O/H) & 8.24 $\pm$ 0.10 & 7.77 $\pm$ 0.06 \\
  12+log(N$^+$/H$^+$) & 7.00 $\pm$ 0.07 & 6.93 $\pm$ 0.04 \\
  log(N/O) & $-$0.81 $\pm$ 0.21 & $-$0.38 $\pm$ 0.12 \\
  12+log(S$^+$/H$^+$) & 5.43 $\pm$ 0.10 & 5.52 $\pm$ 0.07 \\
  12+log(S$^{++}$/H$^+$) & 6.47 $\pm$ 0.12 & 6.27 $\pm$ 0.06 \\
  12+log(S/H) & 6.53 $\pm$ 0.12 & 6.39 $\pm$ 0.06 \\
  log(S/O) & $-$1.69 $\pm$ 0.14 & $-$1.38 $\pm$ 0.09 \\
  12+log(Ar$^{++}$/H$^+$) & 6.11 $\pm$ 0.07 & 5.70 $\pm$ 0.05 \\
  12+log(Cl$^{++}$/H$^+$) & ... & 4.78 $\pm$ 0.13 \\
  He$^+$/H$^+$ (4471) & 0.078 $\pm$ 0.019 & ... \\
  He$^+$/H$^+$ (5876) & 0.077 $\pm$ 0.005 & 0.119 $\pm$ 0.006 \\   
  He$^+$/H$^+$ (6678) & 0.097 $\pm$ 0.006 & 0.083 $\pm$ 0.004 \\
  He$^+$/H$^+$ (7065) & 0.175 $\pm$ 0.019 & 0.079 $\pm$ 0.010 \\
  $<$He$^+$/H$^+$$>$$^{(a)}$ & 0.084 $\pm$ 0.009 & 0.101 $\pm$ 0.018 \\
  He/H & 0.091 $\pm$ 0.009 & 0.123 $\pm$ 0.018 \\
\hline
\end{tabular} 
\end{center}
$^{(a)}$ Mean value excluding He$^+$/H$^+$ (7065 \AA).
\end{table}

Although the excitation degree of M1--78 is rather high and the 
optical nebular \OIII\ lines are strong, the auroral \OIII\ 4363 \AA\ line 
cannot be detected due to the large dust extinction. Fortunately, 
the temperature sensitive \NII\ 5755 \AA\ line was measured with
 a rather good signal-to-noise ratio in both zones and therefore 
we could obtain a direct determination of the electron temperature,
 $T_{\rm e}$, in M1--78. The values of $T_{\rm e}$ are included in 
Table~\ref{table:linesoptical}. It can be seen that the temperature 
is substantially different in the two zones; the NE blob shows 
$T_{\rm e}$ = 13400 K, which is atypically high for a Galactic 
\HII\ region. A previous determination of the temperature was 
obtained by \cite{aller87}, who found $T_{\rm e}$ = 11300 K -- probably
 determined from the \OII\ line ratio -- in agreement with the values we obtain for the 
SW arc. Since the intensity of the \NII\ 5755 \AA\ line 
could suffer from some contribution due to recombination, we 
estimated this contribution using the expression obtained by
 \cite{liu00} and find that it is negligible, approximately 
0.4\% and 0.2\% for the arc and the blob zones, respectively.  

Ionic abundances of N$^+$, O$^+$, O$^{++}$, S$^+$, S$^{++}$, Ar$^{++}$ and 
Cl$^{++}$ were derived from collisionally excited lines by making use of
 the IRAF task IONIC of the package NEBULAR. In the absence of temperature 
determinations for further ions, we considered a one-zone scheme, assuming 
that the temperature obtained from the \NII\ line ratio is valid for the 
rest of the ions in the nebula. We measured several \HeI\ lines in 
our spectra. These lines arise mainly from recombination but they can be 
affected by collisonal excitation and self-absorption effects. We derived the 
He$^+$/H$^+$ ratio using the effective recombination coefficients given 
by \cite{pequignot91} and applying the collisional corrections proposed by 
\cite{kingdon95}. The final adopted value for the He$^+$ abundance is 
the mean of the values obtained for the brightest individual \HeI\ lines but 
excluding \HeI\ 7065 \AA, which suffers from the largest collisional 
effects. All the ionic abundances are included in Table~\ref{table:abundances}. 

Since we do not detect \HeII\ lines, we expect that the contribution 
of O$^{3+}$ should be negligible, and the total O abundance 
was therefore determined by just adding the ionic abundances of O$^+$ and 
O$^{++}$. As  can be seen in Table~\ref{table:abundances}, the O/H 
ratio obtained for the two zones of the nebula 
is very different, the SW arc shows a value consistent with that 
expected by the radial Galactic abundance gradient of e.g. \cite{deharveng00} 
for a galactocentric distance of 12.7 kpc. However, the spectrum of the
 NE blob shows an O/H ratio a factor of 3 
lower than that of the arc. \cite{aller87} also find a relatively low 
O abundance for the object (12+log(O/H) = 8.07). 

The observed  N$^+$/O$^+$ ratio -- that is usually assumed to be equal 
to the N/O ratio due to the similar ionization potential of both ions -- is 
high in both areas of the nebula, especially in the blob, where 
log(N$^+$/O$^+$) = $-$0.38. The expected value of log(N/O) at the distance of the 
nebula should be about $-1.2$ \citep[e.g.][]{shaver83}, 
indicating that the nebula is nitrogen-rich by factors as large as 6 in
the NE blob. The presence of nitrogen enrichment in the object 
has been indicated by several authors, such as \cite{zijlstra90} and \cite{perinotto91}. 

We also estimated the total abundances of He and S by making use 
of an appropriate ionization correction factor, {\it icf}, scheme 
for each element. These factors account for the contribution of the abundances of 
unseen ionization stages of each chemical element. In the case of
 He, we used the {\it icf} proposed by \cite{peimbert92}, 
based on the similarity of the ionization potentials of He$^0$ and 
S$^+$. This {\it icf} seems to work well for relatively high excitation objects 
like M1--78, where only a small fraction of He$^0$ is expected. 
In the case of S, we have to consider some contribution for unseen 
S$^{3+}$ and we assume the correction proposed by \cite{stasinska78}. 
The equations of both {\it icf}s can be found in \cite{garciarojas04}. 
As shown in Table~\ref{table:abundances}, the 
S/H ratios of the two one-dimensional spectra differ by only 40\% 
(0.14 dex) and can be considered rather similar within the errors. 
The fact that the total abundance of S does not show the behavior 
of the O/H ratio seems to support that the difference of O abundance in both zones of the nebula 
could be real and not an artefact produced by, for example, a localized 
abnormally high $T_{\rm e}$. On the other hand, the He/H ratio seems to be 
somewhat higher (34\%) in the NE blob than in the SW arc. 
However, this difference is produced entirely by the different He$^+$/H$^+$ 
ratio obtained from the \HeI\ 5876 \AA\ line in both spectra and is not 
correlated with similar differences in the abundance obtained from the 
other \HeI\ lines. Therefore, we consider the different He/H to be dubious. 

\section{Results from NIR observations}
\label{sect:results:nir}

\subsection{Features of the NIR spectra}

\begin{figure*}
    \begin{minipage}[c]{0.5\textwidth}
      \centering \resizebox{\hsize}{!}{\includegraphics{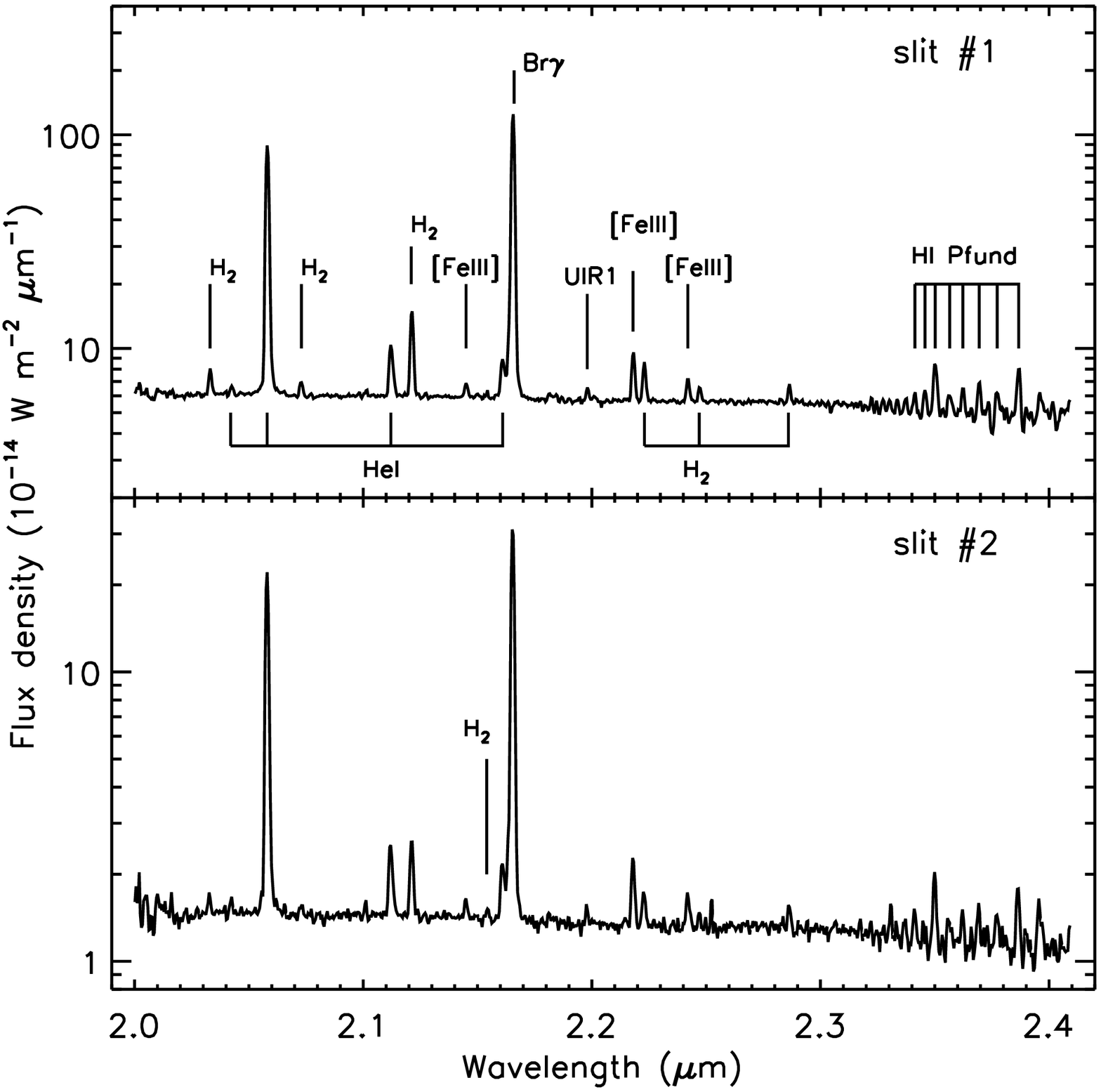}}
    \end{minipage}%
    \begin{minipage}[c]{0.5\textwidth}
      \centering \resizebox{\hsize}{!}{\includegraphics{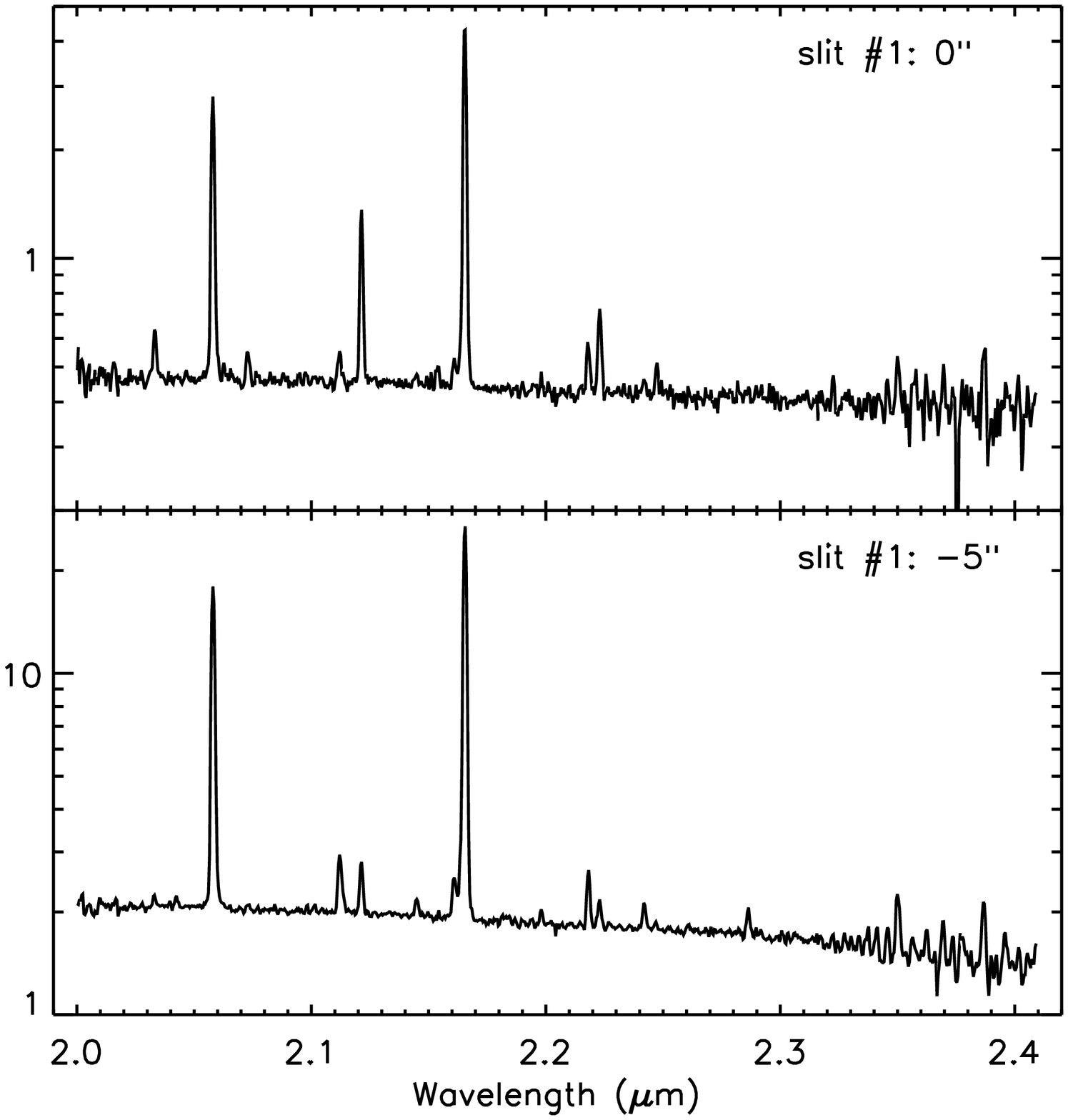}}
    \end{minipage}
     \caption{
       {\bf (Left)}  Spectra of M1--78 extracted across (top) slit \#1 
       between --7\farcs5 and 0\arcsec\  and 
       (bottom) slit \#2 between --2\arcsec\ and 2\arcsec 
       (see Fig.~\ref{fig:im}).
       The observed lines, their identification and
       their line fluxes are listed in Table~\ref{table:lines}.
       {\bf (Right)} Spectra extracted across slit \#1 from 0\farcs75$\times$0\farcs75 regions around the 
       continua detected within the NE blob (top) and the SW arc (bottom). 
       Note the logarithmic scale of the flux axis.
       }
        \label{fig:sp}
   \end{figure*}

Figure~\ref{fig:sp} shows the integrated nebular spectra of M1--78 along
 slits \#1 and \#2. The most obvious features in the spectra are the 
Br$\gamma$ line,  a large set of \HI\ Pfund lines, several \HeI\ emission
 lines and a relatively large number of molecular hydrogen transitions. 
Three forbidden lines of \FeIII\ are also evident. Table~\ref{table:lines} 
lists the observed lines, their identifications and fluxes. The quoted 
uncertainties are those coming from the line fitting procedure. Peak positions, 
equivalent widths and fluxes of the lines were measure by fitting a 
Gaussian profile. A line was defined as being detected if its peak 
intensity exceeds the rms noise level of the local continuum by at least a factor of 3.

\begin{table}
\caption{NIR line fluxes extracted across slit \#1 (between -7\farcs5 and 
0\arcsec) and slit \#2 (between -2\arcsec\ and 2\arcsec). }
\label{table:lines}
\begin{center}
\begin{tabular}{l l r r}
\hline\hline\\[-5pt]
& &  \multicolumn{2}{c}{Line fluxes ($10^{-18}$ W~m$^{-2}$) $^{(a)}$}\\ \cline{3-4} \\[-5pt]
$\lambda$\,(\um) & Identification & \multicolumn{1}{c}{Slit \#1} & 
\multicolumn{1}{c}{Slit \#2}\\[2pt]
\hline\\[-5pt]

  2.033 & H$_2$ 1-0 S(2)              &  28.3 $\pm$ 3.2 &  3.2 $\pm$ 1.0\\
  2.042 & He 6$^3$P-4$^3$S            &   7.7 $\pm$ 1.5 &  2.3 $\pm$ 0.5\\
  2.058 & He 2$^1$P-2$^1$S            &  1406 $\pm$ 39  &  341 $\pm$ 10 \\
  2.073 & H$_2$ 2-1 S(3)              &  15.0 $\pm$ 2.4 & $<3$\\
  2.112 & He 4$^{3,1}$S-3$^{3,1}$P    &  84.7 $\pm$ 4.0 & 20.9 $\pm$ 1.2\\
  2.121 & H$_2$ 1-0 S(1)              & 147.1 $\pm$ 3.5 & 20.5 $\pm$ 0.9\\
  2.145 & \FeIII\ $^3$G$_3$-$^3$H$_4$ &  15.2 $\pm$ 1.4 &  3.5 $\pm$ 0.6\\
  2.154 & H$_2$ 2-1 S(2)              & $<$7          & $<2$\\
  2.161 & He 7$^{3,1}$F-4$^{3,1}$D    &  61.1 $\pm$ 3.3 & 16.0 $\pm$ 1.1\\
  2.165 & Br$\gamma$                  &  1994 $\pm$ 106 &  495 $\pm$ 27 \\
  2.198 & UIR\,1 $^{(b)}$              &  13.1 $\pm$ 2.5 &  2.6 $\pm$ 0.5\\
  2.218 & \FeIII\ $^3$G$_5$-$^3$H$_6$ &  63.4 $\pm$ 1.7 & 15.7 $\pm$ 0.8\\
  2.223 & H$_2$ 1-0 S(0)              &  45.6 $\pm$ 2.2 &  7.1 $\pm$ 1.0\\
  2.242 & \FeIII\ $^3$G$_4$-$^3$H$_4$ &  25.5 $\pm$ 1.3 &  7.6 $\pm$ 1.3\\
  2.247 & H$_2$ 2-1 S(1)              &  14.3 $\pm$ 1.1 & $<5$\\
  2.286 & H$_2$ 3-2 S(2)              &  15.8 $\pm$ 1.8 &  4.0 $\pm$ 1.0\\
  2.340 & \HI\ 31--5                  &  11.4 $\pm$ 3.4 &  3.9 $\pm$ 1.2\\
  2.344 & \HI\ 30--5                  &  16.0 $\pm$ 4.8 & $<8$\\
  2.349 & \HI\ 29--5 $^{(c)}$          &  60.7 $\pm$ 18.2& 12.5 $\pm$ 3.7\\
  2.354 & \HI\ 28--5                  &  16.4 $\pm$ 4.9 &  3.1 $\pm$ 0.9\\
  2.360 & \HI\ 27--5                  &  20.1 $\pm$ 6.0 &  3.8 $\pm$ 1.1\\
  2.367 & \HI\ 26--5                  &  28.0 $\pm$ 8.4 &  5.1 $\pm$ 1.5\\
  2.374 & \HI\ 25--5                  &  19.7 $\pm$ 5.9 &  5.0 $\pm$ 1.5\\
  2.383 & \HI\ 24--5 $^{(d)}$          &  51.4 $\pm$ 15.4& 10.4 $\pm$ 3.1\\

\hline
\end{tabular} 
\end{center}
$^{(a)}$ Errors are at the 1$\sigma$ level and only include the uncertainties 
coming from the fitting procedure. In the case of the \HI\ Pfund series, 
we give a larger error of the order of 30\% because of the uncertainty 
fitting the continuum.
$^{(b)}$ The H$_2$ 3-2 S(3) transition at 2.201 \um\ is observed in 
regions 1 and 4 of slit \#1 (see Table~\ref{tbl:H2}).
$^{(c)}$ This line is probably blended with a transition of \FeIII.
$^{(d)}$ This line is likely contaminated by atmospheric lines.
\end{table}

The validity of our flux calibration can be checked by comparing the 
value we obtain for the Br$\gamma$ line flux with previous spectroscopic
 observations. \cite{hora99} measured a flux of $1110\times10^{-18}$~W~m$^{-2}$ 
with a 1\arcsec$\times$6\arcsec\ slit orientated north--south and 
centered on the brightest infrared portion of the source.  Similarly, 
\cite{lumsden01b} measured $1430\times10^{-18}$~W~m$^{-2}$ integrating 
over a region 1\farcs2$\times$12\farcs6. These values are comparable to 
the flux we observe for Br$\gamma$ and list in Table~\ref{table:lines}.

\subsection{The spatial variation of NIR emission lines}
\label{sect:ir:var}

\begin{figure*}
    \begin{minipage}[c]{0.5\textwidth}
      \centering \resizebox{\hsize}{!}{\includegraphics{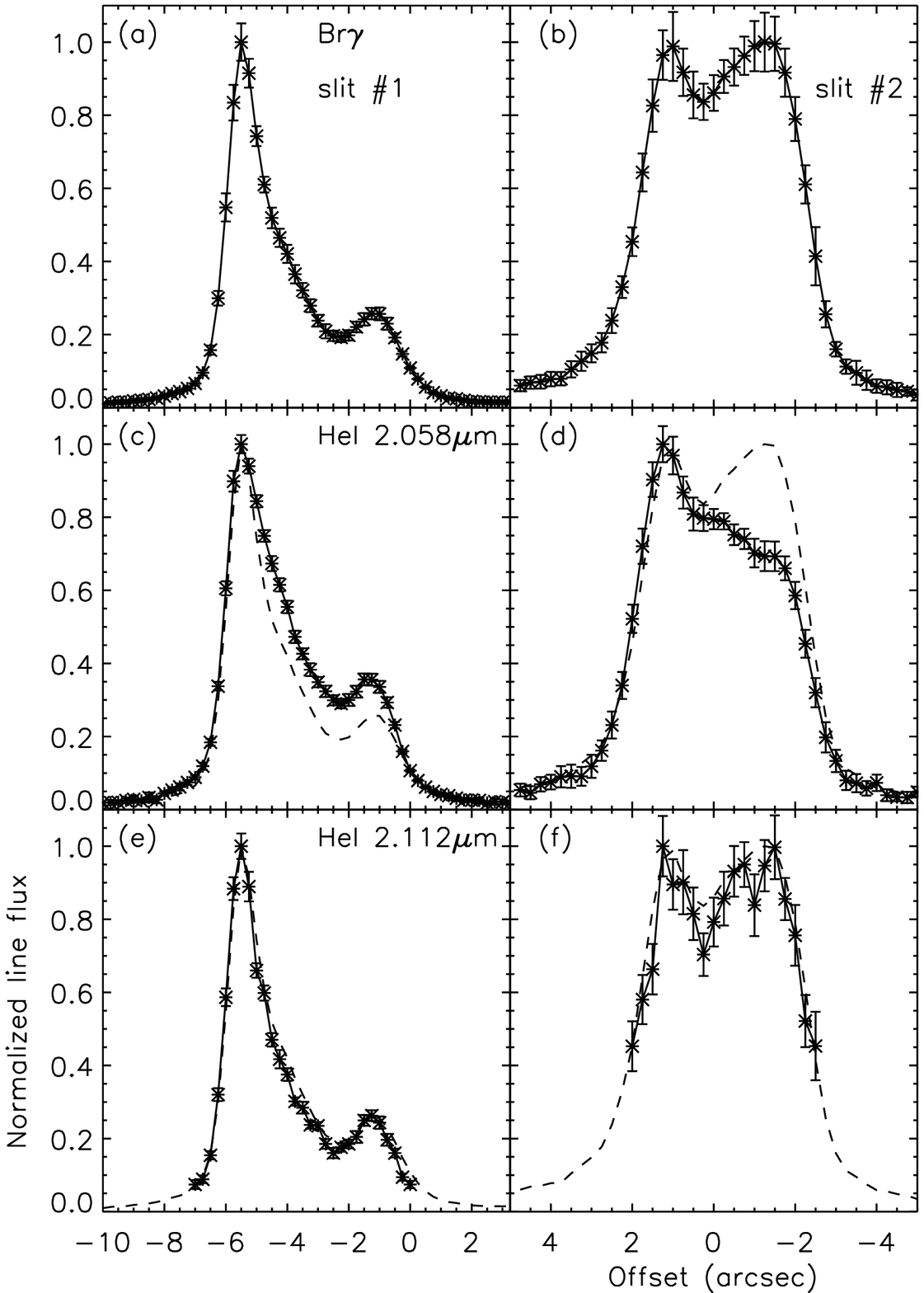}}
    \end{minipage}%
    \begin{minipage}[c]{0.5\textwidth}
      \centering \resizebox{\hsize}{!}{\includegraphics{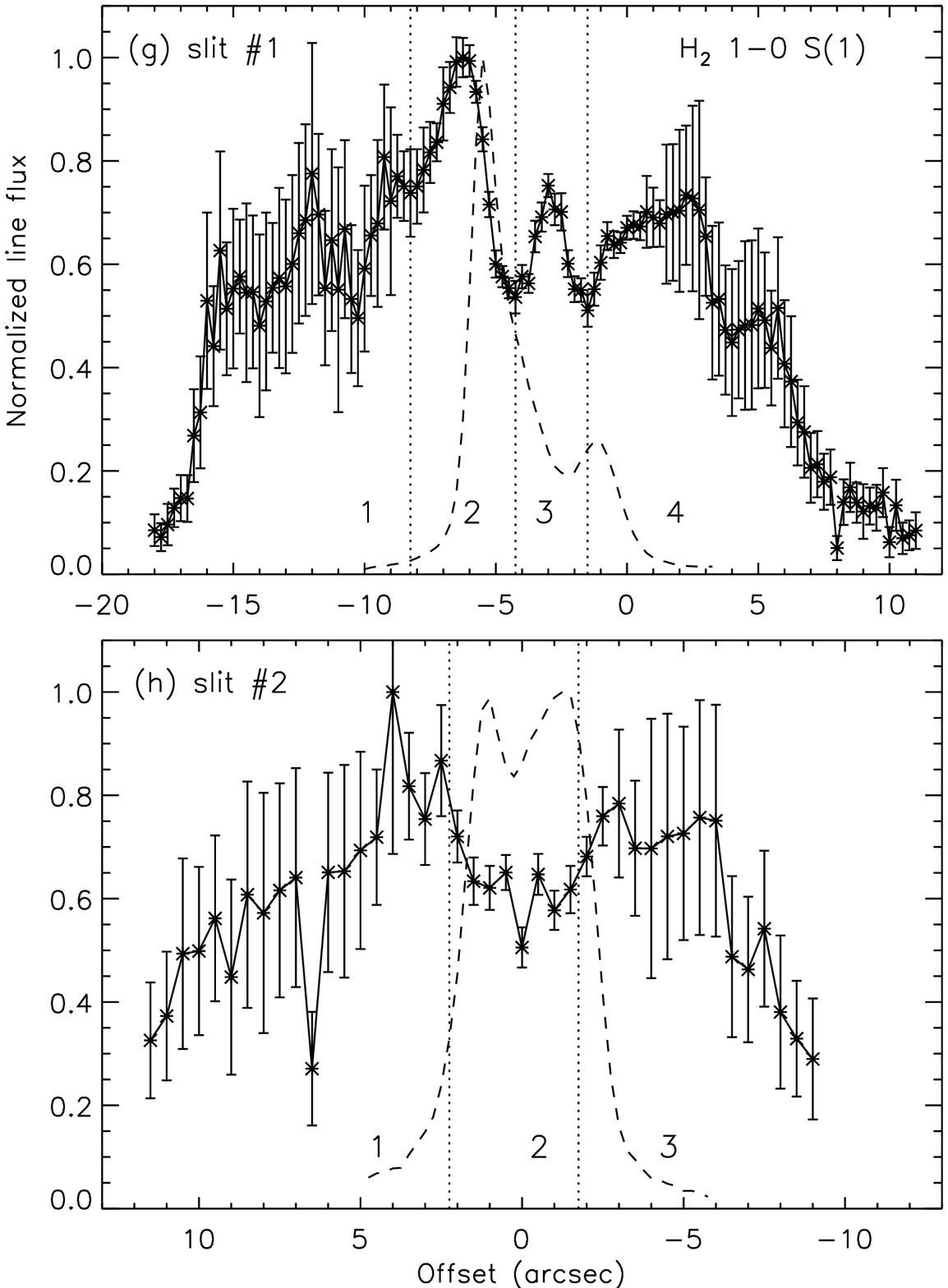}}
    \end{minipage}
     \caption{
       {\bf (Left)} Spatial variation of Br$\gamma$ and the two 
       strongest \HeI\ 
       lines at 2.058 and 2.112~\um\ across slit \#1 (left) and slit \#2 
       (right). The normalized
       \HeI\ intensity distributions are compared to the Br$\gamma$ profile 
       (dashed line). 
       {\bf (Right)} Spatial variation of the H$_2$ 1-0 S(1) transition at 
       2.121~\um\ across slit \#1 (top) and 
       slit \#2 (bottom) compared to the Br$\gamma$ profile (dashed line). 
       The different regions analysed in the text are identified 
       (dotted lines).
       The origin corresponds to the (0\arcsec,0\arcsec) position of
       Fig.~\ref{fig:im}.}
        \label{fig:var12}
   \end{figure*}

Figure~\ref{fig:var12} illustrates the spatial variation of Br$\gamma$, 
the strongest \HeI\ lines (at 2.058 and 2.112~\um) and the H$_2$ 1-0 S(1) 
transition at 2.121~\um. The ionized gas traced by Br$\gamma$ distributes
 along $\sim$7\farcs5 of slit \#1 and $\sim$4\arcsec\ of slit \#2. 
Similarly to the optical observations, the two-dimensional NIR spectrum
 along slit \#1 shows two peaks in the surface brightness distribution of 
the emission lines that correspond to the NE blob and the SW arc. Also
 within these peaks, we detect a featureless and weak continuum, 
specifically at the (0\arcsec,0\arcsec) and (0\arcsec,$-$5\arcsec) locations. 
Spectra extracted from 0\farcs75$\times$0\farcs75 regions around these continua are shown in Fig.~\ref{fig:sp}. 

We investigated the nature of the continua detected in the $K$-band.
 NIR continuum can be produced not only by a stellar photosphere but also
 by: (1) the free--free and free--bound emission from ionized gas, (2) hot 
dust and (3) scattered light, i.e.\ radiation escaping through optically 
thin paths of the nebula and reflected into the line of sight of the observer. 
The contribution by free--free emission to the observed infrared flux density 
can be estimated using equation 7 of \cite{roman04} and the free--free 
Gaunt factors of \cite{hummer88}. For a flux density of 884 mJy at 5\,GHz 
\citep{zijlstra90} and an electron temperature of 10000~K, the expected
 flux density in the $K$-band is about 130\,Jy, which translates into 
$\sim8\times10^{-14}$~W\,m$^{-2}$\,\um$^{-1}$ at 2.2~\um. This is comparable 
to the continuum flux level detected across slit\,\#1 (cf. Fig.~\ref{fig:sp}). 
It seems, therefore, that free--fee emission does indeed dominate the continuum 
observed in the NIR and hides the stellar photospheres detected in the 
optical within the NE blob and the SW arc.

Moving along slit \#1 and from NE to SW, we only start detecting emission
 from Br$\gamma$ at the (0\arcsec,0\arcsec) position, i.e.\ the location
 of the continuum within the NE blob. After this position, the Br$\gamma$ 
line flux slowly increases, reaching a local maximum 1\farcs25 ahead of the
 NE blob and coincident with the weaker blob detected in the acquisition 
image. It then reaches a minimum within the nebula and increases rapidly 
afterwards. This bright portion corresponds to the SW arc structure seen 
in the image and reaches an absolute maximum 5\farcs5 ahead of the NE blob. 

Moving along slit \#2, the Br$\gamma$ emission resembles that of a shell 
structure, with two maxima at the edge of the ionized nebula (at about 
--1\arcsec\ and 2\farcs5) and a decreased emission within the nebula.

The following panels compare the \HeI\ distributions of the 2.058 and 
2.112~\um\ lines to that of the Br$\gamma$. It is noticeable that 
while the 2.112~\um\ \HeI\ line spatial variation traces the 
Br$\gamma$ emission well, this is not the case of the 2.058~\um\ \HeI\ line. 
This difference will be discussed in detail in Sect.~\ref{sect:hei}.

The spatial variation in the H$_2$ 1-0 S(1) transition at 2.121~\um\ is 
also plotted in Fig.~\ref{fig:var12}. The H$_2$ emission extends over 
$\sim$30\arcsec\ across slit \#1 and $\sim$20\arcsec across slit \#2. 
We distinguised four different regions in slit \#1 and three in slit \#2, 
which will be analysed separately in Sect.~\ref{sect:h2}.

\begin{figure}
   \centering \resizebox{\hsize}{!}{\includegraphics{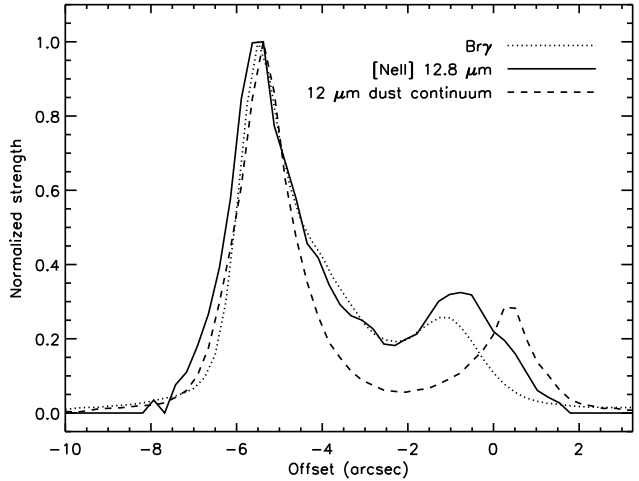}}
     \caption{Spatial variation of the \NeII\ 12.8 \um\ fine-structure line
     (solid line) and the 12 \um\ dust continuum (dashed line) obtained with
     the mid-infrared spectrograph SpectroCam-10 mounted on the 5\,m Hale Telescope 
     at Palomar Observatory. These variations correspond to a 
     1\arcsec$\times$16\farcs4 slit positioned along the same axis of our 
     slit \#1. We show the Br$\gamma$ variation (dotted line) for
     comparison.}
        \label{fig:var3}
   \end{figure}

M1--78 has also been observed using the mid-infrared long-slit spectrograph 
SpectroCam-10 mounted on the 5\,m Hale Telescope at Palomar Observatory 
\cite[cf.][]{peeters:thesis}. These observations were taken using a slit of  
1\arcsec$\times$16\farcs4 slit positioned along the same axis of our slit \#1 
and centered on the brightest spot. Figure~\ref{fig:var3} compares the spatial 
variations of the \NeII\ line at 12.8~\um\ and the flux at 12~\um\ with the 
Br$\gamma$ profile. Both silicate in emission and dust continuum contribute 
to the flux at 12~\um. The variation of the \NeII\ line traces that of 
Br$\gamma$ well, with only a small variation close to the NE blob. The ``dust'' 
profile shows a cavity within the nebula and a clear shell-like structure, 
suggesting that the dust have been removed from within the ionized gas 
volume either through radiation pressure from the star within the NE blob or 
destruction. In contrast, in the arc, the ``dust'' emission concides with 
that of the ionized gas.

\subsection{Unidentified features}

Several unidentified features are also present. Two relatively 
common unidentified features at 2.1981 and 2.2860~\um\ are present 
in the $K$-band spectra of e.g.\ Orion \citep[cf.][]{luhman98} and 
PNe \citep[cf.][]{luhman96,hora96,hora99,lumsden01b}, quite close to 
the 3--2 S(3) and 3--2 S(2) transitions of H$_2$ at 2.2008 and 2.2864~\um,
 respectively, where these are the air wavelengths. This is 
discussed at length by \cite{geballe91}, who examined lists of permitted and 
forbidden atomic lines for wavelength coincidences with these features 
and only found near coincidences with weak lines. Geballe et al.\ claimed 
that the parent ion responsible for these features corresponds to an 
excitation level between 30 and 60~eV, but the higher end of this range 
is inconsistent with the observed presence of these lines in objects 
with effective temperatures $<40000$\,K \citep[cf.][]{lumsden01b}. 
\cite{dinerstein01} claimed an identification for these features as 
[\ion{Kr}{iii}] and [\ion{Se}{iv}] for the 2.198 and 2.286~\um\ lines, 
respectively. However, the abundance of both krypton and selenium in 
the interstellar medium is sufficiently low that this remains a mystery. 

In the integrated spectra shown in Fig.~\ref{fig:sp}, we detect two 
lines at $2.1981\pm0.0001$~\um\ and $2.2864\pm0.0001$~\um. While the 
former line is coincident with the unidentified feature discussed by 
Geballe et al.\ (and designated  UIR\,1), the other line can be 
unquestionably identified as the 3--2 S(2) transition of H$_2$. 
We however note that the 3--2 S(3) transition of H$_2$ (with an 
observed central wavelength at $2.2008\pm0.0001$~\um) is detected in 
regions 1 and 4 of slit \#1, well outside the ionized gas region 
(cf.\ Fig.~\ref{fig:var12}). In these two regions, the UIR\,1 feature 
is not detected. The UIR\,1 feature is not detected either in regions 
1 and 3 of slit \#2. This behavior suggests that the carrier of this 
unidentified UIR\,1 line is probably an ion created by UV photons of 
at least 13.6~eV.

\subsection{Analysis of the \HeI\ recombination lines}
\label{sect:hei}

Calculations of the \HeI\ recombination cascade spectrum have been 
published by different authors, e.g. \cite{smits91, smits96} and 
\cite{benjamin99,benjamin02}. Each study improved the prior treatment 
of physical processes, mainly as the result of improved theoretical 
calculations of various rates. The \HeI\ recombination cascade 
spectrum has been recently revisited by \cite{bauman05} and 
\cite{porter05,porter07} using improved radiative and collisional 
data. They use highly accurate calculations of the $J$-resolved 
transition probabilities that take singlet--triplet mixing into account. 
Calculations of collisionless (low-density limit) Case B \HeI\ 
emissivities are published by \cite{bauman05}. The source of the 
$J$-resolved code used in these calculations is available online 
\citep[see][]{porter07b}.

\begin{figure}
    \centering \resizebox{\hsize}{!}{\includegraphics{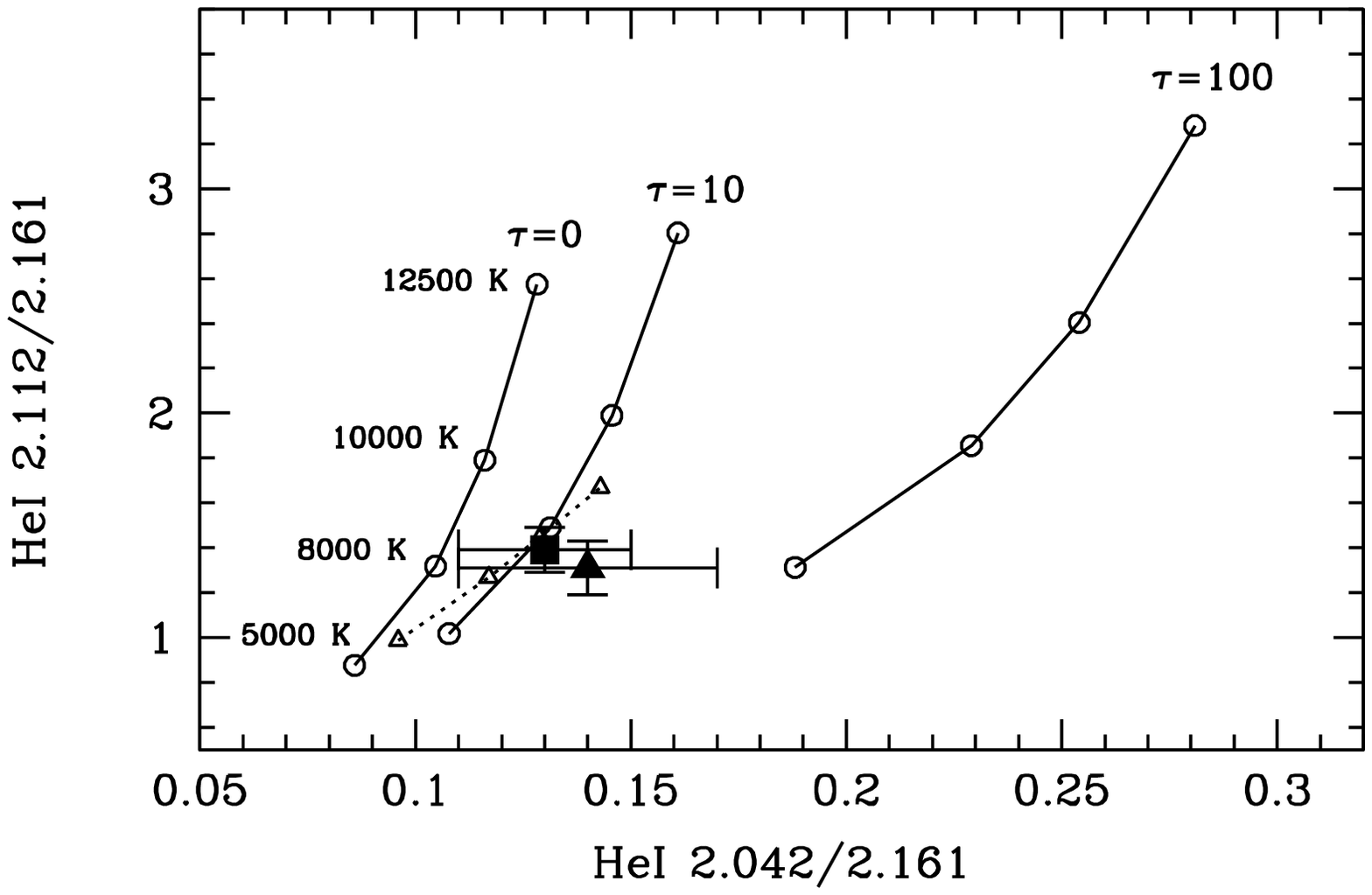}}
     \caption{
      Plot of the \HeI\ 2.112\,4$^{3,1}$S--3$^{3,1}$P/2.161\,7$^{3,1}$F--4$^{3,1}$D ratio against the 
\HeI\ 2.042\,6$^3$P--4$^3$S/2.112\,4$^{3,1}$S--3$^{3,1}$P ratio. Ratios measured for 
slit \#1 are plotted by a black square, and by a black triangle for slit \#2.
	The open triangles represent the theoretical predictions using the 
$J$-resolved code of \cite{bauman05} for $T_{\rm e}=$ 5000~K, 8000~K, 
10000~K and 12500~K. Similarly, the open circles represent the predictions 
of \cite{benjamin99} for the same electron temperatures and three values of $\tau_{3889}$: 0, 10 and 100.
      }  
        \label{fig:he1}
   \end{figure}

Theoretical predictions of several \HeI\ line ratios are shown 
in Fig.~\ref{fig:he1}. We used the calculations of  
\cite{porter07b}, which are only available for the low density limit 
and for $\tau_{3889}$ = 0, and those of \cite{benjamin99}, whose predictions  
consider different values of $n_{\rm e}$ and $\tau_{3889}$. Here, 
$\tau_{3889}$ is the line center optical depth of the 
3$^3$P--2$^3$S 3890 \AA\ line. Since the ratios are rather 
insensitive to $n_{\rm e}$, we only represented the curves 
of \cite{benjamin99} for $n_{\rm e}$ = 10$^4$ cm$^{-3}$ but varying 
$T_{\rm e}$ from 5000 to 12500 K and $\tau_{3889}$ from 0 to 100. 
From the figure, it can be seen that the theoretical predictions of 
\cite{porter07b} with $T_{\rm e}$ between 8000~K and 10000~K 
reproduce the data well, 
whereas the calculations of \cite{benjamin99} give temperatures 
somewhat lower than 8000 K for $\tau_{3889}$ = 10. 
These values of $T_{\rm e}$ -- corresponding to that characteristic
 of the He$^+$ zone of the nebula -- are lower than those obtained 
from the \NII\ line ratios. This is a rather common behavior in 
Galactic HII regions, as can be found in the Orion Nebula \citep{esteban04}, NGC~3576 \citep{garciarojas04} and 
others. The difference between $T_{\rm e}$ obtained from collisionally 
excited and recombination lines could be related to the presence of 
temperature fluctuations in ionized nebulae \citep[e.g.][]{peimbert67}.

Figure~\ref{fig:var12} shows that while the \HeI\ 2.112/Br$\gamma$ ratio 
is fairly constant across the slits, there nevertheless exists a large spatial 
variation in the \HeI\ 2.058\,\um\ line with respect to Br$\gamma$ and 
the other \HeI\ recombination lines. For instance, the \HeI\,2.058/2.112
 ratio varies between 10 and 25, reaching its lowest values at positions 
$<-5$\arcsec for slit \#1 and $>0$\arcsec for slit \#2. The origin of this spatial variation is not clear. To
first order, temperature variations should have similar effects on all 
the lines, so this is probably not the cause. Density variations, on the 
other hand, might be able to produce the observed effect but would 
in principle not be measureable. The most probably cause of this effect 
seems to be optical depth variations across the nebula. 

The  theoretical calculations of \cite{bauman05} fail to reproduce the 
\HeI\ 2.058\,\um\ line. For instance, they predict a \HeI\,2.058/2.112 
ratio around 0.3--0.4. Evidently, the line flux of the \HeI\ 2.058\,\um\ 
line is largely underpredicted. The models of \cite{benjamin99}, on the 
other hand, predict a \HeI\,2.058/2.112 ratio between 20 and 50 for 
large variations of $\tau_{3889}$, density and temperature. Hence, 
models seem incapable of fitting the \HeI\ 2.058\,\um\ line. This is 
because this line is not well defined in the Case B approximation: 
e.g.\ the Bauman et al.\ code assumes that the 2$^1$P to 1$^1$S photons 
are absorbed by some other atom or ion. The \HeI\ 2.058\,\um\ transition 
is determined by the population in the 2$^1$P state. The Case B approximation assumes
that 2$^1$P to 1$^1$S photons do not escape the cloud, but  does that mean that they
are reabsorbed and completely converted to 2$^1$P to 2$^1$S photons, or that
they are destroyed or absorbed by some other atom or ion?  This effect, 
negligible for every other line, has an enormous effect on the 2.058~\um\ line.

\subsection{The \HI\ Pfund lines}   

 The forest of lines redward of about 2.3~\um\ corresponds to the higher 
members of the atomic \HI\ Pfund series. We detected \HI\ pfund 
series lines from Pf24 to Pf31  which allowed us to infer the extinction 
across the $K$-window when compared with the theoretical predictions of 
\cite{hummer87} and \cite{storey95}. These lines are consistent with 
an extinction $A_K< 1$ in agreement with the optical estimate.

\subsection{Forbidden iron lines}

We report the detection of \FeIII\ emission. The $K$-band hosts 
\FeIII\ lines from the multiplet connecting the $^3$G and $^3$H terms 
\citep[e.g.\ figure~6 of][]{bautista98}. Of the four transitions in this 
multiplet with relatively large (magnetic dipole) transition probabilities, 
three are in this wavelength region: 2.145, 2.218 and 2.242~\um. The 
fluxes measured for these lines are listed in Table~\ref{table:lines}. 
Since these lines originate from highly excited levels, their absolute 
emissivities are quite sensitive to the temperature. However, the line 
ratios are primarily sensitive to the density.

\begin{figure}
   \centering \resizebox{\hsize}{!}{\includegraphics{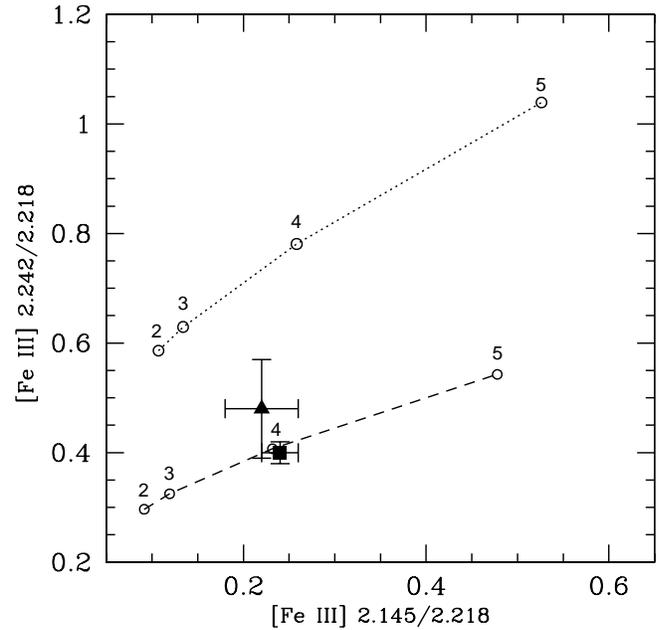}}
     \caption{ $K$-band \FeIII\ line ratios. We show the theoretical 
predictions of \cite{keenan01} (dotted line) and \cite{rodriguez02} 
(dashed line) for densities between log$n_{\rm e}=2$ and log$n_{\rm e}=5$. 
The black square and triangle represent the line ratios measured for slit \#1 and slit \#2, 
respectively.}
        \label{fig:fe3}
   \end{figure}

We compared our observed \FeIII\ line ratios with theoretical predictions. 
We used the calculations of \cite{keenan01} and the more complete 
ones provided by M.~Rodr\'{i}guez \citep{rodriguez02}. These last 
calculations use a 33-level model atom where all collision strengths are those calculated by \cite{zhang97}, 
the transition probabilities those recommended by \cite{froesefischer98} 
\citep[and those from][for the transitions 
not considered by Froese Fischer \& Rubin]{garstang58}, and the level 
energies have been taken from the NIST database. We considered densities  between 
10$^{2}$ and 10$^{5}$~cm$^{-3}$. In Fig.~\ref{fig:fe3} we can see that 
our data points are consistent with the predictions of 
Rodr\'\i guez, giving densities on the order of those determined from 
the optical forbidden lines. However, the predictions of Keenan et al.\ fail 
to fit the observed 2.242/2.218~\um\ ratios; this is also the case for 
the theoretical predictions of \cite{bautista98}, who overpredict the 
observed 2.242/2.218~\um\ ratios by factors  between $\sim$1.5 and 2.5.

\begin{table*}
\caption{Air wavelengths, upper energy levels, Eistein A coefficients, degeneracies and line fluxes of the H$_2$ lines detected in M1--78.
We also give significant line flux ratios and results such as the 
ortho-to-para ratio ($\phi$) and the excitation temperature.}
\label{table:h2}
\begin{center}
\begin{tabular}{c@{\hspace{0pt}}c@{\hspace{4pt}}r@{\hspace{4pt}}c@{\hspace{4pt}}r@{\hspace{5pt}}r@{\hspace{5pt}}r@{\hspace{7pt}}r@{\hspace{7pt}}r@{\hspace{7pt}}c@{\hspace{5pt}}r@{\hspace{7pt}}r@{\hspace{7pt}}r}
\hline\hline\\[-5pt]
Transition & $\lambda_{\rm j}$ & $E_{\rm j}$ & $A_{\rm j}$ & $g_{\rm j}$ & \multicolumn{4}{c}{Slit 1 $^{(a)}$} & & \multicolumn{3}{c}{Slit 2 $^{(a)}$}\\[1pt]
\cline{6-9} \cline{11-13} \\[-5pt]
& (\um) & (K) & (10$^{-7}$ s) & & Region 1 & Region 2 & Region 3 & Region 4 & & Region 1 & Region 2 & Region 3
\\[2pt]
\hline\\[-5pt]
1-0 S(2) & 2.0332 &  7584 & 3.98 &  9 &  35.3 $\pm$ 2.6  & 18.5 $\pm$ 1.9 & 10.4 $\pm$ 1.1 &  26.3 $\pm$ 1.1 & &  8.0 $\pm$ 1.4 &  2.9 $\pm$ 0.7 &  5.3 $\pm$ 1.0\\
2-1 S(3) & 2.0729 & 13890 & 5.77 & 33 &  16.7 $\pm$ 1.9  &  9.9 $\pm$ 1.1 &  4.6 $\pm$ 0.6 &  22.3 $\pm$ 1.1 & & 5.9 $\pm$ 1.1 & $<3$           &  3.1 $\pm$ 0.7\\
1-0 S(1) & 2.1212 &  6956 & 3.47 & 21 & 148.6 $\pm$ 10.4 & 92.3 $\pm$ 1.4 & 48.5 $\pm$ 1.2 & 150.8 $\pm$ 3.6 & & 44.5 $\pm$ 2.3 & 19.1 $\pm$ 0.8 & 30.1 $\pm$ 1.9\\
2-1 S(2) & 2.1536 & 13152 & 5.60 &  9 & $< 3.5$          & $<4.2$         & $<2.5$         &  10.7 $\pm$ 1.1 & & 3.6 $\pm$ 0.9 & $<2$           & $<3$          \\
3-2 S(3) & 2.2008 & 19086 & 5.63 & 33 &  5.7 $\pm$ 1.4   & $<4.2$         & $<2.2$         &   8.5 $\pm$ 2.3 & & $<3$           & $<3$           & $<2$          \\
1-0 S(0) & 2.2229 &  6471 & 2.53 &  5 & 33.8 $\pm$ 3.8   & 28.5 $\pm$ 2.2 & 16.6 $\pm$ 0.8 &  38.3 $\pm$ 1.7 & & 11.6 $\pm$ 1.5 &  6.9 $\pm$ 1.0 & $<5$          \\
2-1 S(1) & 2.2471 & 12550 & 4.98 & 21 & 15.1 $\pm$ 2.2   &  9.3 $\pm$ 1.1 &  4.2 $\pm$ 0.7 &  16.3 $\pm$ 1.6 & & 5.0 $\pm$ 1.1 & $<4$           &  4.0 $\pm$ 1.1\\
3-2 S(2) & 2.2864 & 18386 & 5.63 &  9 & $< 3.3$          & 11.6 $\pm$ 1.3 &  4.0 $\pm$ 0.6 & $< 4.5$         & & $<3$           &  4.4 $\pm$ 1.0 &  2.3 $\pm$ 1.0\\[2pt]
\hline\\[-5pt]

1-0 S(1)/2-1 S(1)   & & & & & 9.8 $\pm$ 1.6  & 9.9 $\pm$ 1.2     & 11.5 $\pm$ 1.9    & 9.2 $\pm$ 0.9  & & 8.6 $\pm$ 1.9 & $>4$           & 7.5 $\pm$ 2.1 \\
1-0 S(1)/3-2 S(3)   & & & & & 26.1 $\pm$ 6.6 & $> 22$            & $> 22$            & 17.6 $\pm$ 4.8 & & $>17$         & $>7$           & $>15$         \\
$\phi_1$            & & & & & 3.5 $\pm$ 0.3  & 3.1 $\pm$ 0.2     & 2.9 $\pm$ 0.2     & 3.7 $\pm$ 0.1  & & 3.6 $\pm$ 0.4 & 3.3 $\pm$ 0.4  & $^{(b)}$       \\
$\phi_2$            & & & & & $\vartriangle$ & $\vartriangle$    & $\vartriangle$    & 2.0 $\pm$ 0.2  & & 1.7 $\pm$ 0.4 & $^{(c)}$ & $^{(c)}$\\
$T_{\rm ex}$ (K) $^{(e)}$  & & & & & 980 $\pm$ 165  & 720 $\pm$ 55      & 690 $\pm$ 20      & 720 $\pm$ 150  & & 750 $\pm$ 175 & 570 $\pm$ 55   & 590 $\pm$ 120 \\[2pt]
$T_{\rm ex}$ (K) $^{(f)}$ & & & & & 2255 $\pm$ 230 & 2100 $\pm$ 260    & 1910 $\pm$ 330    & 2465 $\pm$ 390 & & 2380 $\pm$ 330& $^{(d)}$     & 2265 $\pm$ 420\\[2pt]

\hline
\end{tabular}
\end{center}
\label{tbl:H2}
$^{(a)}$ Line fluxes are given in units of 10$^{-18}$~W\,m$^{-2}$.
$^{(b)}$ The 1-0 S(0) line was not been measured.
$^{(c)}$ Only one or two $v=2$ lines were detected.
$^{(d)}$ There are only upper limits for the $v=2$ transitions.
$^{(e)}$ Excitation temperature determined from the $v=1$ states only.
$^{(f)}$ Excitation temperature determined from the $v=1$ and $v=2$ states.
\end{table*}

\subsection{H$_2$ excitation}
\label{sect:h2}

The NIR emission from H$_2$ can be produced by either thermal 
emission in shock fronts or fluorescence excitation by non-ionizing 
UV photons in the Lyman--Werner band (912--1108\,\AA). These mechanisms 
can be distinguished since they preferentially populate different 
levels producing different spectra. For shock-excited H$_2$ the lower 
energy levels are typically populated as for a gas in local thermal 
equilibrium (LTE) with a characterisitic excitation temperature around 
1000\,K. In contrast, for radiatively excited gas the population 
follow a non-LTE distribution characterized by high excitation temperatures 
around 10\,000\,K and lines from high vibrational levels ($v \gg 1$) may be 
detected. 

Non-thermal excitation mechanisms readily excite the $v=2$ and higher 
vibrational states, whereby collisional transitions preferentially 
de-excite the $v=2$ level in favor of the $v=1$ level. Hence, the 
H$_2$ 1--0 S(1)/2--1 S(1) ratio has on occasion been used to distinguish 
between these two processes. However, in dense clouds, FUV irradiation 
can heat the gas to temperatures $\sim$1000\,K leading to collisionally 
excited and thermalized 1--0 line emission, while direct FUV pumping 
maintains non-thermal fluorescent populations in higher-lying vibrational 
levels \citep[e.g.][]{sternberg89,draine96}.
Hence, for low densities ($n_{\rm H} \sim 10^3$ cm$^{-3}$), the 1--0 
S(1)/2--1 S(1) ratio retains a pure fluorescence value around 2. But 
for high density clouds ($n_{\rm H} \gtrsim 10^3-10^4$ cm$^{-3}$) and 
an intense incident FUV radiation field with $G_0 \gtrsim 10^4$ (where 
$G_0$ is the incident FUV photon flux between 6 and 13.6~eV measured 
in units of the local interstellar radiation field), the 1--0 S(1)/2--1 
S(1) ratio will rapidily approach a thermal value of about 10 
\citep[e.g.][]{hollenbach95}, which has often been interpreted to indicate shock emission.
Conversely, the 1--0 S(1)/3--2 S(3) ratio retains a value of 
approximately 8 in a dense photodissociation region (PDR) when $G_0 \sim 10^3$ and only for 
very high densities and a very intense FUV radiation field ($n_{\rm H} 
\gtrsim 10^5$ cm$^{-3}$ and $G_0 \gtrsim 10^4$) will this ratio 
approach a thermal value of 10--100 \citep{hollenbach95}. Therefore, 
the combination of the 1--0 S(1)/2--1 S(1) and 1--0 S(1)/3--2 S(3) 
ratios should in principle serve as an indicator of the excitation 
mechanism of the H$_2$.

We distinguished four different regions along slit \#1 and 
three along slit \#2 with sufficient signal-to-noise (cf.\ Fig.~\ref{fig:var12}). 
The line fluxes measured for all these regions are listed in 
Table~\ref{table:h2}. This table also lists the values for the 
1--0 S(1)/2--1 S(1) and 1--0 S(1)/3--2 S(3) ratios. Across slit \#1, 
the 1--0 S(1)/2--1 S(1) remains practically constant within the errors 
with an average value around $9.7\pm0.6$. The 1--0 S(1)/3--2 S(3) 
ratio is in general higher than 20 for all these regions. A similar 
scenario occurs for the three regions along slit \#2. Both ratios suggest 
thermal emission. However, in the case of a dense PDR with 
$n_{\rm H} \gtrsim 10^5$ cm$^{-3}$ and a very intense FUV radiation 
field with $G_0 \gtrsim 10^4$, these ratios can be explained by a 
non-thermal excitation mechanism. We can now estimate if the PDR in 
M1--78 can effectively be characterized by such high values of $G_0$ and $n_{\rm H}$.

$G_0$ can be derived by assuming that all the UV photons are 
absorbed in a spherical shell of the size of the PDR and re-emitted 
in the infrared continuum \citep[cf.][]{bernard05}. Normalizing $G_0$ 
to the average interstellar field \citep[$1.6\times10^{-6}$~W m$^{-2}$;][]{habing68}, the expression is:

\begin{equation}
 G_0 = {{4F_{\rm IR}} \over {1.6\times10^{-6} \times 2.35\times10^{-11} \theta^2}}~,
\end{equation}

\noindent
where $\theta$ is the diameter of the PDR in arcseconds and 
$F_{\rm IR}$ is the observed infrared flux in W~m$^{-2}$. We
 use a FIR flux of $\sim~7\times10^{-12}$ W~m$^{-2}$ determined 
by fitting a blackbody through the IRAS fluxes at 25, 60 and 
100 \um\ \citep{peeters:catalogue}. We obtain a $G_0$ around 
$(1-2)\times10^3$ considering a size of 20\arcsec--30\arcsec\ 
(see Fig.~\ref{fig:var12}). We note, however, that $G_0$ might be 
substantially larger if the filling factor of the PDR gas is small.

The total density can be determined, for instance, using 
fine-structure lines emitted in the PDR. The \OI\ 63\,\um/\CII 
158\,\um\ ratio is a good tracer for densities in between $10^3$ and 
$10^6$ cm$^{-3}$ because these two lines have different critical 
densities. These two lines have been measured by ISO within an 
aperture of $\sim80$\arcsec\ \citep{peeters:catalogue}. With a ratio of $7.5\pm0.6$, the models by \cite{kaufman99} 
indicate a density of $10^4-10^5$ cm$^{-3}$ when $G_0 \gtrsim 10^3$.

Therefore, it seems reasonable to assume the case of a very dense PDR and a strong radiation field.
This analysis, however, is not conclusive and we will use two additional diagnostics to clarify the
 origin of the H$_2$ emission: the ortho-to-para ratio and the excitation diagrams.

\subsubsection{The ortho-to-para ratio}

An additional diagnostic is provided by the ortho-to-para ratio. At 
the moment of formation, molecular hydrogen can join one of two 
denominations: ortho (aligned nuclear spins) or para (opposed nuclear spins). 
The ortho-to-para ratio $\phi$, based on the statistical weights of the nuclear spins, is 3:1. 
In outflow regions where shock excitation is the primary emission 
mechanism, observations of vibrational emission lines typically reveal 
ortho-to-para ratios for vibrationally excited states that are comparable 
to 3 \citep[e.g.][]{smith97}, as is expected if the gas behind such shocks is in LTE.
However, as discussed at length by \cite{sternberg99}, the ortho-to-para 
ratio can take values other than the intrinsic 3 if the excitation is 
caused by UV fluorescence, since the transitions are pumped by optically 
thick UV transitions in which the optical depth depends on whether the 
molecule is in an ortho or para state. Indeed, ratios in the range 
1.5--2.2 have been measured in PDRs \citep[e.g.][]{ramsay93,chrysostomou93,shupe98,lumsden01}.

From the 1--0 S(0), 1--0 S(1) and 1--0 S(2) line fluxes, we measured
 the ortho-to-para ratio for the $v=1$ states ($\phi_1$) following 
equation 9 of \cite{smith97}. In Table~\ref{table:h2} we list the $v=1$ 
ortho-to-para ratio for each region. These are consistent with a ratio 
of 3 suggesting a thermalization of the $v=1$ states. Of course, they do 
not tell the whole story. In a few cases, a value of the ortho-to-para 
ratio for the $v=2$ states ($\phi_2$) was derived from the 2--1 S(1), 
2--1 S(2) and 2--1 S(3) transitions using equation 1 of \cite{davis03}.
 We obtained ratios around 2, typical of PDRs and in agreement with the
ratios derived by \cite{lumsden01b}.

  \begin{figure*}
   \centering \resizebox{\hsize}{!}{\includegraphics{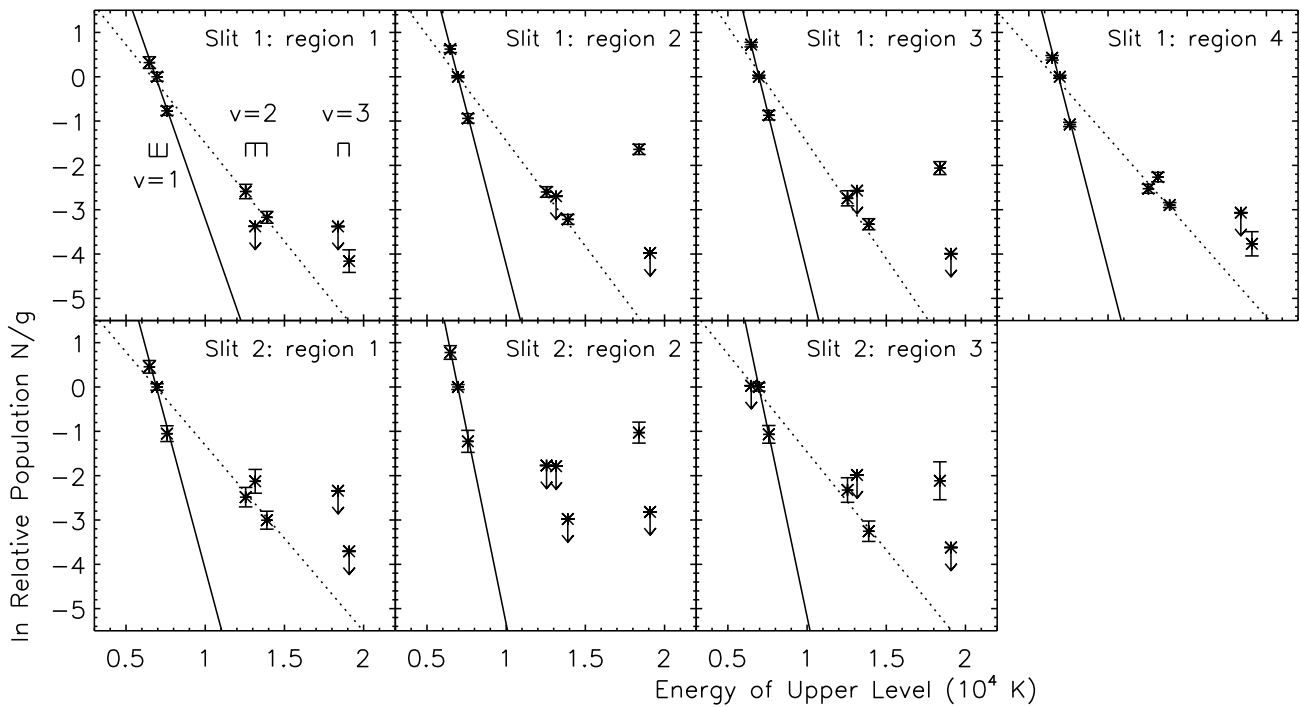}}
     \caption{H$_2$ excitation diagrams for each region. The lines 
     represents the best fitting purely thermal single-temperature models. The 
     solid line is the fit to the $v=1$ level only and the dotted line is 
     the fit to all the data except the $v=3$ level.}
        \label{fig:h2}
   \end{figure*}

\subsubsection{$H_2$ excitation diagrams}

The simplest way of characterizing the molecular hydrogen emission, however,
is to plot the observed column density against the energy of the upper level.
The measured intensity, $I$, of a given H$_2$ line can be used to
 calculate the column density of the upper excitation level of the
 transition, which, for optically thin emission, is given by:

\begin{equation}
 N_j = { {4 \pi \lambda_j I} \over {A_j h c} }~,
\end{equation}

\noindent
where $\lambda_j$ is the rest wavelength and $A_j$ is the 
Einstein A-coefficient taken from \cite{turner77}. If collisional
 de-excitation is assumed to dominate, the H$_2$ will be in LTE and 
the energy levels will be populated in a Boltzmann distribution. The 
relative column densities of any two excitation levels can thus be 
expressed in terms of an excitation temperature $T_{\rm ex}$:

\begin{equation}
 { N_i \over N_j} = {g_i \over g_j} {\rm exp} \left [ { {-(E_i -E_j)} \over {k T_{\rm ex}} } \right ]~,
\end{equation}

\noindent
where $g_j$ is the degeneracy, $E_j$ is the energy of the upper level 
taken from \cite{dabrowski84} and $k$ is the Boltzmann's constant. 
The values of $\lambda_j$, $E_j$, $A_j$ and $g_j$ for the lines 
detected in the spectra are shown in Table~\ref{table:h2}. Clearly, 
plotting the logarithm of the ratio of the column densities divided 
by the appropiate statistical weights gives the excitation temperature 
from the slope of the data.

In Fig.~\ref{fig:h2} we show plots of log$N_j/g_j$ versus the energy 
for each region. We normalized the population distributions relative 
to that inferred from the 1--0 S(1) line. For thermalized populations at 
a fixed gas temperature, the log$N/g$ points should lie on a straight 
line in these excitation diagrams. Cleary, such single-component models 
do not fit the data. While the $v=1$ levels appear to be thermalized at 
$T \sim 600-1000$\,K (solid lines), the inclusion of the $v=2$ levels 
raises the best-fitting temperature to $T \sim 2000-2500$\,K (dotted 
lines). \cite{lumsden01b} find excitation temperatures of 2400\,K at 
the position of the nebula and slightly higher (around 3500\,K) to the
 north and south. 

However, even at these high temperatures, the $v=3$ levels remain 
underpopulated and the excitation temperature inferred from the 
relative populations in the $v=2$ and $v=3$ levels exceeds 4000\,K. 
At such gas temperatures, the molecules would be rapidly dissociated, 
suggesting that a non-thermal excitation mechanims is responsible for 
the excitation of the $v=2$ and $v=3$ levels.

The results of these diagrams indicate again that the observed H$_2$ 
excitation is produced in a dense PDR rather than in shocks. In fact, 
a high vibrational excitation temperature, such as the 4000\,K inferred 
from the relative $v=2$ and $v=3$ populations, is a basic characteristic
 of FUV-pumped fluorescent emission \citep[e.g.][]{burton90}. 
Furthermore, and as pointed out by \cite{davies03}, an additional 
signature of fluorescent excitation appears to be present in the data. 
For those regions in which the 2--1 S(1), S(2) and S(3) lines are detected 
(region 4 in slit \#1 and region 1 in slit \#2), the resulting $N/g$ 
values for the $j$=3, 4 and 5 levels do not appear to be thermalized. Instead, 
the $j=4$ para-level is shifted above the adjacent $j=3$ and $j=5$ ortho-levels. 
The ortho-to-para ratio for the $v=2$ transitions appears to be out of 
equilibrium and suppressed below the value of 3 obtained in LTE. This is 
precisely what is expected in the case of FUV pumping, where the populations 
of the vibrationally pumped ortho-H$_2$ is suppressed relative to those of 
the para-H$_2$ \citep{sternberg99}. Indeed, we find values as low as 2 for 
the $v=2$ ortho-to-para ratio.

\section{Discussion: the nature of M1--78}
\label{sect:discussion}

\cite{zijlstra90} and \cite{perinotto91} reported that the chemical 
composition of M1--78 is enriched in nitrogen with respect to 
oxygen when compared with the typical abundance pattern of \HII\ 
regions. Our optical observations, much deeper than previous ones, confirm this 
enrichment but also indicate that it seems 
to be stronger in the NE blob. The blob coincides with a very 
faint stellar emission that can be seen in our red optical  
(as discussed in Sect.~\ref{sect:ir:var}, the NIR continuum is on the 
other hand dominated by free--free nebular emission). Therefore, it seems 
that a process of localized chemical enrichment is taking place to the 
NE of M1--78, and that the star we see at the blob is its most 
probable source. At the blob, the nitrogen enrichment is correlated with
 a defficiency in the O abundance and a -- dubiou -- He enrichment. This 
abundance pattern is typical of nebular ejecta around evolved massive stars such
as WR \citep[e.g.][]{esteban92} and LBV stars \citep[e.g.][]{lamers01}. 
The chemical composition of these nebulae reflects the product of the 
CNO nucleosynthesis in the H-burning zones of the parent post-main-sequence massive star. 

In order to explore the likeliness of  stellar ejecta in M1--78, 
we derived its ionized mass from the radio flux of 884~mJy at 
5 GHZ obtained by \cite{zijlstra90}. Considering a distance of 8.9 
kpc and a diameter of 8\farcs0, we obtain $n_{\rm e}$(rms) $\sim$ 6200\,cm$^{-3}$ and 
M(\HII) $\sim$ 3\,M$_\odot$. The fact that the $n_{\rm e}$(rms) 
coincides with the $n_{\rm e}$ obtained from the optical forbidden lines indicates 
that the filling factor of the gas is very close to 1. The ionized mass 
derived for M1--78 is on the order of those found for ejecta nebulae 
around WR and LBV stars, typically about a few solar masses 
\citep[e.g.][]{esteban92,lamers01}. However, our nominal value of 3\,M$_\odot$ 
should be considered as an upper limit of the mass of the ejecta. 

The spatial variations of the physical conditions and chemical abundances and the 
presence of more than one possible ionizing source indicate that M1--78 
is better described as a combination of a compact \HII\ region + ejecta. 
In this sense, this combined nature can explain the negative result of 
\cite{hutsemekers97}, who doubts the classification of M1--78 as an ejecta nebula 
due to its very large dust mass obtained from the far-infrared emission 
measured by the IRAS satellite. We have to consider that the ejecta nebulae that 
\cite{hutsemekers97} uses for his comparisons are isolated and are not related 
to a nearby star-forming region as M1--78 presumably is. 
Further indirect evidence of an ejecta nebula in M1--78 
comes from the aforementioned detection of strong \NiqII\ emission 
in its spectrum. In fact, \cite{lucy95} indicates that circumstellar 
nebulae around LBV stars are anomalous \NiqII\ emitters, as has been found in the nebula around 
P~Cygni \citep{johnson92} and two objects in the LMC \citep{stahl86}. 
However, the high excitation and $T_{\rm e}$ of M1--78 do not support 
its identification as a LBV nebula. Typically, these objects show 
$T_e$ lower than 10000 K \citep[e.g.][]{lamers01,nota95}, and very 
faint or even unobservable \OIII\ \citep{nota95}. Obviously, this difficulty 
can be obviated if there is another hotter ionizing source in the nebula apart of the hypothetical LBV star itself. 
In contrast, ring nebulae around WR stars show a wide range of nebular 
excitations and $T_{\rm e}$, basically depending on the spectral classification 
of the ionizing WR star. In fact, $T_{\rm e}$ between 12000 and 14000 K 
have been found in nitrogen-rich nebulae ionized by early WN and WO stars as NGC~2359, 
S~308 or G~2.4+1.4 \citep{esteban90,esteban92}.

Another remarkable feature of M1--78 is the expansion velocity of the 
ionized gas of about 25 km s$^{-1}$ measured 
by \cite{gussie89}. Although this fact was considered by \cite{gussie95} 
as an indication of the PN nature of the object, such 
supersonic expansion velocities are also common in WR ring nebulae 
\citep[e.g.][]{rosado86} as well as in LBV nebulae 
\citep[e.g.][]{nota95}. 

A close analogue of M1--78 is N82 \citep{heydari82}, a compact \HII\ 
region in the Large Magellanic Cloud. This object shows nitrogen 
enrichment and oxygen defficiency. The ejecta seem to be produced by 
a WC9 star and there is one additional ionizing O star in the nebula 
(\citealt{moffat91}, however, argued against this classification and 
suggested that the presence of a WNL or Of star in N82 is more likely). 
The linear size, ionized mass and $n_{\rm e}$ of N82 
(0.4 pc, 7 M$_\odot$ and 5500 cm$^{-3}$, respectively) are very similar 
to those of M1--78 but the higher excitation and $T_{\rm e}$ 
of this Galactic object would indicate that the progenitor should be a 
WC or WN star of an earlier -- hotter -- subclass.  

We investigate now the possibility that a second star, specifically 
an O star, might be present in M1--78, e.g.\ at the position of the
 SW arc where a second stellar continuum is detected in the optical.
An estimate of the spectral type of the ionizing star in \HII\ regions 
based on the \HeI\ 2.112/Br$\gamma$ ratio has been proposed by 
\cite{hanson02}. This method relies on a measure of the ionization fraction 
of helium in the \HII\ region. For very hot stars 
($T_{\rm eff} \gtrsim 40\,000$\,K), helium is ionized throughout the 
entire nebula producing a constant relative strength of the recombination 
lines of helium and hydrogen. For a $T_{\rm eff}$ below this value, 
the helium emission compared to hydrogen quickly decreases as the
 \HII\ region region becomes predominantly neutral helium. Using 
the predictions of \cite{bauman05}, we determined the ratio 
of \HeI\ 2.112~\um\ to Br$\gamma$ when the helium is fully ionized 
in an \HII\ region. Assuming $n({\rm He})/n({\rm H})=0.10$ and an 
electron temperature of 10000\,K, these models predict a ratio of 
0.040, in perfect agreement with the observed ratio in both slits 
($0.042\pm0.002$, see Table~\ref{table:lines}). Assuming for now 
that a single O star is responsible for the ionization of the
 nebula, this would imply that the ionizing star must be an early 
O star with $T_{\rm eff} \gtrsim 40\,000$\,K. This would correspond 
to a spectral type earlier than O5.5\,V, as inferred from the new 
calibration of O star parameters published by \cite{martins05} 
and based on state-of-the-art stellar models.

We do indeed find a \HeI\,2.112/Br$\gamma$ ratio  of $0.043\pm0.003$ at 
the position of the SW arc. However, this ratio is lower at the position 
of the NE blob; specifically, \HeI\,2.112/Br$\gamma=0.029\pm0.005$ in 
the 0\farcs75$\times$0\farcs75 region around the stellar continuum within the NE blob.  
This implies either a low electron temperature ($< 5000$\,K) in the 
NE blob (which is not the case, see Section~\ref{sect:opresults}) or 
the presence of a colder star that is not able to fully ionize He. 
It therefore appears that the WR star that we suspect to be hiding within the 
NE blob is not solely responsible for the ionization of the nebula.

It seems reasonable, then, to assume that a early O star is present in 
the SW arc of M1--78. The NIR spectrum of such a massive star will 
be characterized by emission of \NIII\ at 2.115~\um\ but the detection 
of this line at intermediate spectral resolutions is difficult since 
it blends with the \HeI\ line at 2.112~\um. Moreover, it is not surprising 
that we have not detected direct signs of its presence, since the NIR 
emission we detect is merely nebular. Detection of stellar features 
in the optical spectrum is also difficult because of the high visual extinction.

The spectral type of the ionizing stars in \HII\ regions can be also 
determined from radio continuum measurements. \cite{zijlstra90} measured 
884~mJy at 5~GHz. Assuming that the nebula is dust-free, optically thin, 
homogeneous and ionization-bounded, this density flux gives a minimum value for log$Q_0$ of 48.80~dex, 
where $Q_0$ is the rate of Lyman continuum photons required to maintain the ionization of the nebula.
This value of $Q_0$ would correspond to a star with a spectral type 
earlier than O6.5\,V star, in perfect agreement with the previous estimate.

\section{Conclusions}
\label{sect:conclusions}

There is considerable controversy surrounding the nature of M1--78, a compact object 
located beyond both the Local arm and the Perseus arm. It was first classified 
as a PN and is nowadays generally considered to be a compact \HII\ 
region. To investigate the nature of M1--78 further, we obtained 
long-slit, intermediate-resolution, optical spectroscopy with the ISIS spectrograph 
at the WHT. As a complement, we obtained long-slit, intermediate-resolution, 
NIR spectra using LIRIS, the NIR imager/spectrographer also installed at the WHT.

M1--78 is characterized by two main morphological zones, a bright arc to the 
SW and a blob of emission to the NE, which correspond to two peaks 
in the surface brightness distribution of the optical and NIR emission lines. 
In addition, each peak contains a featureless, very weak and unresolved stellar 
continuum that was detected in the optical.

The electron densities calculated from the \SII\ 6717/6731 line ratio are found 
to be very high and similar in both zones, 6700 and  6400 cm$^{-3}$ for the arc 
and the blob. These are consistent with the densities determined from the NIR 
\FeIII\ line ratios. The electron temperature determined from the optical nebular 
\NII\ lines, on the other hand, is substantially different in the two zones, 10900~K 
in the SW arc and 13400~K in the NE blob. This last temperature is atypically 
high for a Galactic \HII\ region. The temperature determined from the NIR \HeI\ 
lines is somewhat lower, but this difference between $T_e$ obtained from collisionally 
excited and recombination lines could be related to the presence of temperature 
fluctuations. We find as well local differences in the extinction. The spectrum 
of the NE blob shows a somewhat larger reddening coefficient than the SW arc, 
indicating a larger amount of internal dust in this zone that produces a higher 
$A_V$ of about 9~mag. 

The most important result, however, is the confirmation of a nitrogen enrichment 
in M1--78. This enrichment 
is stronger at the location of the NE blob. It seems that a process of localized 
chemical enrichment is taking place in the NE of M1--78. This N enrichment in the blob is correlated 
with a defficiency in the O abundance and a -- dubious -- He enrichment. Such abundance 
pattern is typical of ejecta nebulae around evolved massive stars such as WR stars 
and LBV stars. The high excitation and electron temperature found in M1--78 suggest 
that it is more likely to be a WR star: ring nebulae around WR stars show a wide 
range of nebular excitation and $T_e$ while LBV nebulae usually show $T_e$ lower 
than 10000~K and very faint or even unobservable \OIII. 

The spatial variations in the physical conditions and chemical abundances and the 
presence of more than one possible ionizing source indicates that M1--78 is better 
described as a combination of a compact \HII\ region $+$ ejecta. In fact, the 
\HeI\,2112\,\um/Br$\gamma$ line ratio, which is commonly used to estimate the 
spectral type of ionizing stars in \HII\ regions, indicates a hot 
($T_{eff} \gtrsim 40000$~K) O star in the SW arc. In conlusion, M1--78 seems to be 
ionized by a WR $+$ hot O star system. Further observations able to observe directly 
the phosphere of these stars will be necessary to confirm this result.

Finally, we detect H$_2$ emission wich extends over a large ($\sim30$\arcsec) 
area around the ionized nebula. The analysis of the NIR H$_2$ lines, which includes 
estimates of the ortho-to-para ratios and excitation diagrams, indicates that the 
excitation mechanism is UV fluorescence. We do not find evidence of shocks.

\begin{acknowledgements}
We thank M\'onica Rodr\'\i guez for her help with the theoretical predictions 
of the \FeIII\ line ratios, and Ryan Porter for his input regarding the theoretical 
calculations of the \HeI\ recombination lines.
 During this work, NLMH has been supported by a Juan de la Cierva fellowship from 
the Spanish Ministerio de Ciencia y Tecnolog\'{\i}a (MCyT). This work has also been 
partially funded by the Spanish MCyT under project AYA2004-07466.
\end{acknowledgements}


\end{document}